\pdfoutput=1
\RequirePackage{ifpdf}
\ifpdf % We are running pdfTeX in pdf mode
\documentclass[pdftex]{sigma}
\else
\documentclass{sigma}
\fi

\numberwithin{equation}{section}
\newtheorem{Proposition}{Proposition}[section]

\theoremstyle{definition}{
\newtheorem*{rem}{Remark}}

\usepackage{rotating,booktabs}

\begin{document}

\allowdisplaybreaks

\newcommand{\arXivNumber}{1911.01180}

\renewcommand{\PaperNumber}{015}

\FirstPageHeading

\ShortArticleName{Classical Superintegrable Systems in a Magnetic Field}

\ArticleName{Classical Superintegrable Systems in a Magnetic Field\\ that Separate in Cartesian Coordinates}

\Author{Antonella MARCHESIELLO~$^\dag$ and Libor \v{S}NOBL~$^\ddag$}

\AuthorNameForHeading{A.~Marchesiello and L.~\v{S}nobl}

\Address{$^\dag$~Czech Technical University in Prague, Faculty of Information Technology,\\
\hphantom{$^\dag$}~Department of Applied Mathematics, Th\'{a}kurova 9, 160 00 Prague 6, Czech Republic}
\EmailD{\href{mailto:marchant@fit.cvut.cz}{marchant@fit.cvut.cz}}

\Address{$^\ddag$~Czech Technical University in Prague, Faculty of Nuclear Sciences and Physical Engineering,\\
\hphantom{$^\ddag$}~Department of Physics, B\v rehov\'a 7, 115 19 Prague 1, Czech Republic}
\EmailD{\href{mailto:Libor.Snobl@fjfi.cvut.cz}{Libor.Snobl@fjfi.cvut.cz}}

\ArticleDates{Received November 05, 2019, in final form March 06, 2020; Published online March 12, 2020}

\Abstract{We consider superintegrability in classical mechanics in the presence of magnetic fields. We focus on three-dimensional systems which are separable in Cartesian coordinates. We construct all possible minimally and maximally superintegrable systems in this class with additional integrals quadratic in the momenta. Together with the results of our previous paper [\textit{J.~Phys. A: Math. Theor.} \textbf{50} (2017), 245202, 24~pages], where one of the additional integrals was by assumption linear, we conclude the classification of three-dimensional quadratically minimally and maximally superintegrable systems separable in Cartesian coordinates. We also describe two particular methods for constructing superintegrable systems with higher-order integrals.}


\Classification{37J35; 78A25}	

\section{Introduction}
In this paper we investigate superintegrability of three-dimensional systems that separate in Cartesian coordinates in the presence of a magnetic field. We say that a mechanical system is superintegrable if it is Liouville integrable and possesses additional independent integrals of motion. Depending on their number we distinguish minimal superintegrability when only one additional integral is present, and maximal superintegrability when the number of additional integrals is the maximal possible, i.e., equal to the number of degrees of freedom minus one. (In three spatial dimensions there is no other possibility.)

The study of superintegrability with magnetic fields was initiated in~\cite{DoGraRaWin} and subsequently followed in both two spatial dimensions~\cite{BeWin,CharHuWin,Pucacco,PuRos} and three spatial dimensions~\cite{BS, MS,MS2,MSW,MSW2}; relativistic version of the problem was recently considered too, cf.~\cite{HeiIld}. Separability of three-dimensional systems with magnetic fields was considered in the papers~\cite{BeChaRas, ShaBaMe}. Particular planar two-body systems, e.g., Coulomb, in perpendicular constant magnetic field were also studied from the point of solvability and superintegrability, see, e.g., \cite{TurEsc1,Taut1,Taut2,TurEsc2}.

It turns out that the presence of magnetic field significantly increases the complexity of both calculations and structure of these systems. E.g., contrary to the case without magnetic field separability in orthogonal coordinates and integrability with integrals at most quadratic in the momenta are no longer equivalent, namely separability is stronger and implies the existence of at least one integral linear in the momenta. Similarly, the explicit construction of superintegrable systems and their classification become much harder when magnetic fields are present.

In the present paper we attempt to approach the problem from a different viewpoint. We exploit the fact that in certain situations the three-dimensional system can be rewritten as effectively a two-dimensional one without magnetic field, thus generalizing the principal idea of~\cite{MS2}. In other cases we show that the existence of a quadratic integral necessarily implies the existence of an integral in a particular simpler form, which makes our calculations tractable. When the results of the present paper and~\cite{MS, MSW} are viewed together, they provide an exhaustive list of three-dimensional quadratically minimally and maximally superintegrable systems with magnetic fields separable in Cartesian coordinates.

We shall investigate the superintegrability of the system defined on the phase space $\mathbb{R}^6$, with the canonical coordinates $(\vec x,\vec p)$, by
\begin{gather}\label{Hamiltonian}
H(\vec x,\vec p)=\frac12\big(\big(p_1^A\big)^2+ \big(p_2^A\big)^2+\big(p_3^A\big)^2\big)+ W(\vec x),
\end{gather}
where $W(\vec x)$ denotes the so called electrostatic or effective potential, $p_j^A$ are the covariant expressions for the momenta
\begin{gather}\label{covarP}
 p_j^A=p_j+ A_j(\vec x),\qquad j=1,2,3,
 \end{gather}
and $A_j(\vec x)$ are the components of the vector potential. The magnetic field $\vec B(\vec x)$ is related to~$\vec A(\vec x)$ through
\begin{gather*}
\vec B(\vec x)=\nabla\times\vec A(\vec x).
\end{gather*}
Newtonian equations of motion and thus also the physical dynamics are gauge invariant, i.e., depend only on $B(\vec x)$ and $\nabla W(\vec x)$. However, in the Hamiltonian formulation gauge transformations can be seen as canonical transformations (cf.\ \cite[Problem~11.25]{KotSer}), namely they alter the Hamiltonian, the corresponding Hamilton's equations of motion and the Hamilton--Jacobi equation in a prescribed way. Separation of variables in the Hamilton--Jacobi equation is related to a~specific choice of the coordinate system and is not preserved under canonical transformations~-- on the contrary, one looks for a suitable canonical transformation such that the system becomes separable after it. Since we are interested in systems that separate in Cartesian coordinates, we find it preferable to work in a suitably chosen fixed gauge adapted to the separation.

Furthermore, we will sometimes use canonical transformations to reduce to cyclic coordinates corresponding to integrals. Also in this perspective, it is helpful to fix an appropriate gauge. However, the final results, in particular the superintegrable systems found shall be given in the gauge covariant form, so to express them in the most general way.

In gauge dependent form the Hamiltonian \eqref{Hamiltonian} reads
\begin{gather}\label{Hamiltonian_gauge}
H(\vec x,\vec p)=\frac12\big(p_1^2+ p_2^2+p_3^2\big)+ A_1(\vec x) p_1 + A_2(\vec x)p_2 + A_3(\vec x) p_3+ V(\vec x),
\end{gather}
where the gauge dependent ``scalar'' potential $V(\vec x)$, i.e., the momentum-free term in~\eqref{Hamiltonian_gauge}, is related to the gauge invariant electrostatic potential $W(\vec x)$ via
\begin{gather*}
V(\vec x)=W(\vec x)+ \frac12 \big|\vec A(\vec x)\big|^2.
\end{gather*}
There are only two cases in which the system~\eqref{Hamiltonian_gauge} separates in Cartesian coordinates \cite{BeChaRas,ShaBaMe}, up to a canonical permutation of the variables. Let us write them in both gauge dependent and gauge covariant form:

\textbf{Case I}
\begin{gather}\label{Sep1}
 V(\vec x)=V_1(x_1)+ V_2(x_2),\qquad \vec A(\vec x)=(0,0,u_1 (x_2)-u_2 (x_1)),
\end{gather}
therefore
\begin{gather}\label{Omega_sep1}
\vec B(\vec x)=(u_1' (x_2), u_2' (x_1),0),\qquad W(\vec x)=V_1(x_1)+ V_2(x_2)-\frac12(u_1(x_2)-u_2(x_1))^2.
\end{gather}

\textbf{Case II}
 \begin{gather}\label{Sep2}
 V(\vec x)=V_1(x_1),\qquad \vec A(\vec x)=(0,u_3 (x_1),-u_2(x_1)),
 \end{gather}
 thus
 \begin{gather}\label{Omega_sep2}
 \vec B(\vec x)=(0, u_2' (x_1),u_3'(x_1)),\qquad W(\vec x)=V_1(x_1)-\frac12\big(u_3(x_1)^2+u_2(x_1)^2\big).
 \end{gather}

In these two cases the system admits two Cartesian-type integrals, related to the separation of variables:
\begin{gather}
X_1=\big(p_1^A\big)^2-2(u_2(x_1)(p_3^A-u_1(x_2)+u_2(x_1))-V_1(x_1))=p_1^2- 2 (u_2(x_1) p_3- V_1(x_1)),\nonumber\\
X_2=\big(p_2^A\big)^2+2(u_1(x_2)(p_3^A-u_1(x_2)+u_2(x_1))+V_2(x_2))=p_2^2+ 2 (u_1(x_2) p_3+V_2(x_2))\!\!\!\!\label{CartInt1}
\end{gather}
for~\eqref{Sep1} and
\begin{gather}
X_1=p_2^A-u_3(x_1)= p_2,\qquad X_2=p_3^A- u_2(x_1)=p_3\label{CartInt2}
\end{gather}
for \eqref{Sep2}.
\begin{rem}
$X_0=p_3^A - u_1(x_2)+ u_2 (x_1)=p_3$ is another integral of~\eqref{Omega_sep1}, though dependent on the Hamiltonian and \eqref{CartInt1}.
\end{rem}

Minimal superintegrability due to the existence of another first-order integral has been studied in \cite{MS, MSW}. Here we investigate the conditions for the existence of an additional integral of order at least two for the systems~\eqref{Omega_sep1},~\eqref{Omega_sep2}. We give an exhaustive list of systems for which an additional second-order integral exists, and are able to answer the question on the existence of higher-order integrals in special cases.

Sections \ref{sec:minimally_ext} and \ref{sec:maximally_higher} present two propositions for finding out whether certain classes of systems are superintegrable by reducing to a two-dimensional (2D) problem without magnetic field. In this way we also construct families of systems with higher-order integrals. Next, in Section~\ref{sec:second_ord_int} we address the problem of second-order superintegrability. The determining equations for second-order integrals are given in gauge covariant form, together with their compatibility conditions. In Section~\ref{sec:necessary} we give a necessary condition for second-order superintegrability, which is used in Sections~\ref{sec:case1} and \ref{sec:case2} to simplify the structure of the integral for the classes \eqref{Omega_sep1}, \eqref{Omega_sep2}, respectively. With these simplifications at hand, the determining equations for the integral can be solved. In Section~\ref{Conclusion1} we list the superintegrable systems so found; their explicit derivation is rather technical and tedious and we review it in Appendices~\ref{appendix:case1},~\ref{appendix:case1-part2} and~\ref{appendix:case2}. The special case in which the magnetic field is constant and the functions $V_j$ in \eqref{Omega_sep1} and \eqref{Omega_sep2} are at most quadratic polynomials is studied in Section~\ref{sec:pol2}. Finally, in Section~\ref{Conclusion2} we discuss the approaches to construction of higher-order integrals.

\section[Minimal superintegrability for Case~I when all the integrals commute with one linear momentum]{Minimal superintegrability for Case~I\\ when all the integrals commute with one linear momentum}\label{sec:minimally_ext}

Let us consider the natural Hamiltonian systems on the phase space $(x_1,x_2,p_1,p_2)$, for $\kappa$ $\in\mathbb{R}$, $\kappa\neq0$
\begin{gather}\label{2dof}
\mathcal H^{\kappa}_0(x_1,x_2,p_1,p_2)=\frac12\big(p_1^2+p_2^2\big)+ \kappa(u_1(x_2)- u_2(x_1))+ V_1(x_1)+ V_2(x_2).
\end{gather}
For the sake of clarity let us refer here to the Hamiltonian of Case I as to $\mathcal H$.
 Since $p_3$ is an integral of motion for~\eqref{Sep1}, by setting $p_3=\kappa$, $\mathcal H_0^{\kappa}= \mathcal H(x,y,z,p_1,p_2,\kappa)-\frac{1}{2}\kappa^2 $. Both systems have a pair of second-order integrals corresponding to separation: $X_j$ as in~\eqref{CartInt1} for $\mathcal H$ and clearly
\begin{gather*}
\mathcal I_j^{\kappa}= X_j(x_1,x_2,p_1,p_2,\kappa),\qquad j=1,2
\end{gather*}
for \eqref{2dof}.

If \eqref{Sep2} possesses any additional integral $X_3$ independent of the variable $x_3$, then
$\mathcal I_3^{\kappa}(x_1,x_2,p_1,\allowbreak p_2)=X_3(x_1,x_2,p_1,p_2,\kappa)$ would be an integral for~\eqref{2dof}. And vice versa, any additional integral~$\mathcal I_3^{\kappa}$ of~\eqref{2dof}, would correspond to an integral $X_3$ of~\eqref{Sep2}, obtained by simply replacing~$\kappa$ by~$p_3$, i.e., $X_3(x_1,x_2,x_3,p_1,p_2,p_3)=\mathcal I_3^{p_3}(x_1,x_2,p_1,p_2)$. Indeed,
\begin{gather*}
\{ \mathcal H, X_3 \}=\sum_{i=1}^2\left(\frac{\partial \mathcal H_0^{p_3} }{\partial {x_i}}\frac{\partial \mathcal I_3^{p_3}}{\partial {p_i}}-\frac{\partial\mathcal I_3^{p_3} }{\partial {x_i}}\frac{\partial \mathcal H_0^{p_3} }{\partial {p_i}}\right) + \frac12 \big\{p_3^2, X_3 \big\}=0,
\end{gather*}
where $\{\,,\,\}$ is the Poisson bracket on the phase space $\mathbb{R}^6$.
The right hand side of the equality is zero since both $\mathcal H$ and $X_3$ do not depend on $x_3$ and $\mathcal I_3^{p_3}$ is an integral of $\mathcal H_{0}^{p_3}$. Thus, we arrive at the following immediate conclusion
\begin{Proposition}\label{lemma_minimally}
Let us consider the Hamiltonian system defined by \eqref{Hamiltonian} on the phase space $(x_1,x_2,x_3,p_1,p_2,p_3)$ with magnetic field and effective potential as in \eqref{Omega_sep1}. Such system admits an additional independent integral $I_3$ such that $\{I_3,p_3\}=0$ if and only if \eqref{2dof} is superintegrable on the phase space $(x_1,x_2,p_1,p_2)$.
\end{Proposition}

Therefore all the systems of the form \eqref{Omega_sep1} that are minimally superintegrable, with an additional integral independent of Cartesian coordinate, can be deduced from 2D natural superintegrable systems of the form \eqref{2dof}. And vice versa, every superintegrable system in two degrees of freedom can be extended to a minimally superintegrable system in three degrees of freedom with magnetic field. Superintegrable systems of the form \eqref{2dof} have been widely studied. In particular they have been completely classified for integrals up to third order \cite{MiPoWin}. Concerning higher-order integrals, many examples are known, including the harmonic oscillator and the caged oscillator~\cite{EvaVe,RTW}, and a wide class of so called exotic potentials~\cite{EscLVWin1,EscWinYur,MarSajWin}.

\subsection{Example: extension of 2D second-order superintegrable systems}
Table \ref{table:2Dquad} contains all three-dimensional systems that can be proven to be (at least) minimally quadratically superintegrable by applying Proposition \ref{lemma_minimally} to 2D superintegrable systems that separate in Cartesian coordinates and have integrals at most quadratic. The list of 2D systems is taken from \cite{MiPoWin}, from which we consider only the systems on real phase space. To obtain the most general family of systems (and recalling that the Hamiltonian must depend linearly on $\kappa$), we renamed all the parameters as $c_j=a_j \kappa + b_j$, $a_j$ not all vanishing, then set $p_3=\kappa$ and applied Proposition \ref{lemma_minimally}. The third integral, leading to superintegrability, can then be found from the integral $\mathcal I_{3}$ of the 2D system, by substituting $c_j=a_j p_3 + b_j$. Since the dependence on the constants $c_j$ is linear, the order of the so obtained integral remains quadratic.
\begin{sidewaystable}
\centering
\caption{3D (at least) minimally quadratically superintegrable extensions of 2D quadratically superintegrable systems that separate in Cartesian coordinates. For the reader's convenience, we give the Hamiltonian expressed in the gauge choice~\eqref{Sep1}, but also the functions $u_j$ and $V_j$ that allow to find the magnetic field $\vec B$ and potential $W$ as in the more general gauge invariant form~\eqref{Omega_sep1}. In the integrals, $L_3$ denotes angular momentum on the plane, $L_3=x_1 p_2- x_2 p_1$. }\label{table:2Dquad}
\vspace{2mm}

\begin{tabular}[htbf]{|cc|c|}
 \hline
 & 2D system and its third integral & 3D system \\
 \hline
 $\mathcal E_1$: & $\begin{array}{c}
 \mathcal H_0= \frac12\big(p_1^2+p_2^2\big)+ c_1\big(x_1^2+ x_2^2\big)+ \frac{c_2}{x_1^2}+ \frac{c_3}{x_2^2}\\
 \mathcal I_3= L_3^2 +2\big(c_2\frac{x_2^2}{x_1^2}+ c_3\frac{x_1^2}{x_2^2}\big)\end{array}$ & $\begin{array}{c}
 \mathcal H=\frac12\big(p_1^2+p_2^2+p_3^2\big)+\big(a_1\big(x_1^2+ x_2^2\big)+ \frac{a_2}{x_1^2}+ \frac{a_3}{x_2^2}\big) p_3+ b_1\big(x_1^2+ x_2^2\big)+ \frac{b_2}{x_1^2}+ \frac{b_3}{x_2^2} \tsep{2pt}\\
 u_1(x_2)= a_1 x_2^2+\frac{a_3}{x_2^2},\qquad u_2(x_1)= -a_1 x_1^2-\frac{a_2}{x_1^2} \\ V_1(x_1)= b_1 x_1^2+\frac{b_2}{x_1^2},\qquad V_2(x_2)= b_1 x_2^2+\frac{b_3}{x_2^2} \end{array}$ \bsep{2pt}\\
 \hline
 $\mathcal E_2$: & $\begin{array}{c}
 \mathcal H_0=\frac12\big(p_1^2+p_2^2\big)+c_1\big(4 x_1^2+ x_2^2\big) + c_2 x_1 + \frac{c_3}{x_2^2}\\
 \mathcal I_3= p_2 L_3-x_2^2\big( 2 c_1 x_1+ \frac{c_2}{2}\big)+ 2c_3\frac{x_1}{x_2^2}
 \end{array}$ & $\begin{array}{c}
 \mathcal H=\frac12\big(p_1^2+p_2^2+ p_3^2\big)+\big(a_1\big(4 x_1^2+ x_2^2\big) + a_2 x_1 + \frac{a_3}{x_2^2}\big)p_3+ b_1\big(4 x_1^2+ x_2^2\big) + b_2 x_1 + \frac{b_3}{x_2^2} \tsep{2pt}\\
 u_1(x_2)=a_1 x_2^2 +\frac{a_3}{x_2^2},\qquad u_2(x_1)=-4 a_1 x_1^2-a_2 x_1 \\ V_1(x_1)=4 b_1 x_1^2+b_2 x_1,\qquad V_2(x_2) = b_1 x_2^2+\frac{b_3}{x_2^2} \end{array}$ \bsep{2pt}\\
 \hline
 $\mathcal E_3$: & $\begin{array}{c}
 \mathcal H_0=\frac12\big(p_1^2+p_2^2\big)+c_1\big(x_1^2 + x_2^2\big)+ c_2 x_1+ c_3 x_2\\
 \mathcal I_3=p_1 p_2 + 2 c_1 x_1 x_2 + c_2 x_2+ c_3 x_1
 \end{array}$ & $ \begin{array}{c}
 \mathcal H=\frac12\big(p_1^2+p_2^2+ p_3^2\big)+ \big(a_1\big(x_1^2 + x_2^2\big)+ a_2 x_1+ a_3 x_2\big) p_3 +b_1\big(x_1^2 + x_2^2\big)+ b_2 x_1+ b_3 x_2 \tsep{2pt}\\
 u_1(x_2)=a_1 x_2^2 + a_3 x_2,\qquad u_2(x_1)=-a_1 x_1^2-a_2 x_1 \\ V_1(x_1)=b_1 x_1^2+ b_2 x_1,\qquad V_2 (x_2)=b_1 x_2^2 + b_3 x_2
 \end{array}$ \bsep{2pt}\\
 \hline
\end{tabular}

\end{sidewaystable}

\subsection[Example: a family of higher-order superintegrable systems from the 2D caged oscillator]{Example: a family of higher-order superintegrable systems\\ from the 2D caged oscillator}\label{sec:minimally_higher}

Let us consider the two-dimensional caged anisotropic oscillator
\begin{gather}\label{2Dcage}
\mathcal {H}_0=\frac12\big(p_1^2+p_2^2\big)+\omega\big(\ell^2 x_1^2+m^2 x_2^2\big)+ \frac{\alpha}{x_1^2}+\frac{\beta}{x_2^2}
\end{gather}
for $\omega\in\mathbb{R}\setminus\{0\}$, $\ell$, $m$ nonvanishing integers and $\alpha,\beta\in\mathbb{R}$. The system is well known to be superintegrable if $\frac{\ell}{m}$ rational~\cite{EvaVe,RTW}. A first straightforward extension to a~3D superintegrable system is given by
\begin{gather}\label{extcage1}
\mathcal {H}=\frac12\big(p_1^2+p_2^2+ p_3^2\big)+\big(\ell^2 x_1^2+m^2 x_2^2\big)p_3+ \frac{\alpha}{x_1^2}+\frac{\beta}{x_2^2},
\end{gather}
that can be transformed into \eqref{2Dcage} by simply reducing $p_3=\omega$.

A more general extension can be constructed as in the previous example. Let us set
\begin{gather}\label{newparameters2}
\ell^2= \ell_1 \kappa +\ell_2,\qquad \alpha=\alpha_1\kappa+\alpha_2, \qquad
m^2=m_1 \kappa+m_2 ,\qquad \beta=\beta_1\kappa+\beta_2.
\end{gather}
The system \eqref{2Dcage} can then be seen as the 2D reduction of
\begin{gather}
\mathcal {H} = \frac12\big(p_1^2+p_2^2+p_3^2\big)+\left(\omega\big(\ell_1 x_1^2+m_1 x_2^2\big)+\frac{\alpha_1}{x_1^2}+\frac{\beta_1}{x_2^2}\right)p_3 \nonumber \\
\hphantom{\mathcal {H} =}{} + \omega\big(\ell_2 x_1^2+m_2 x_2^2\big) + \frac{\alpha_2}{x_1^2}+\frac{\beta_2}{x_2^2},\label{extcage2}
\end{gather}
by substituting $p_3=\kappa$. We obtain in this way the three-dimensional integrable system \eqref{extcage2} that becomes superintegrable when the frequency ratio of \eqref{2Dcage} (where \eqref{newparameters2} has to be taken into account) is a rational number, i.e., when
\begin{gather}\label{ratio1}
\frac{\ell_1 p_3+\ell_2}{m_1 p_3+ m_2 }=\frac{\ell^2}{m^2},\qquad \frac{\ell}{m}\in\mathbb{Q},
\end{gather}
for every possible value of the phase space variable~$p_3$. Equivalently, \eqref{ratio1} can be written as
\begin{gather*}%\label{ratio2}
\big(m^2 \ell_1-\ell^2 m_1\big)p_3+ m^2\ell_2-m_2\ell^2=0, \qquad \frac{\ell}{m}\in\mathbb{Q}.
\end{gather*}
The above equation contains a polynomial in $p_3$ that must be identically zero. This is possible only when the coefficient of each power of~$p_3$ vanishes. Namely, when
\begin{gather}\label{ratio_cond}
\frac{\ell_1}{m_1}=\frac{\ell_2}{m_2}=\frac{\ell^2}{m^2},\qquad \frac{\ell}{m}\in\mathbb{Q}.
\end{gather}
Thus, the family of systems~\eqref{extcage2} is superintegrable if and only its parameters satisfy \eqref{ratio_cond} (and in that case also \eqref{2Dcage} is superintegrable). For $\ell_j=m_j=0$ for some $j$ (not both $j=1,2$), the previous condition reduces to
\begin{gather*}
 \frac{\ell_j}{m_j}=\frac{\ell^2}{m^2},\qquad \frac{\ell}{m}\in\mathbb{Q}.
\end{gather*}
For $\alpha_1=\beta_1=\ell_2=m_2=0$ we have the simpler system~\eqref{extcage1}.
The case $\alpha_j=\beta_j=0$, $\ell_j=m_j=\pm1$, $j=1,2$ was studied in~\cite{MS} and it is shown there to be quadratically minimally superintegrable, with the fourth independent integral (besides the two Cartesian ones) inherited from the 2D caged oscillator, of first order. In the more general case \eqref{extcage2}, the order of the fourth integral can be arbitrarily high, depending on the value of $\frac{\ell}{m}$. Notice that all the systems in Table~\ref{table:2Dquad} are contained in the family \eqref{extcage2}, except the systems $\mathcal{E}_2$ and $\mathcal{E}_3$ for the special case $a_1=b_1=0$ (i.e., $c_1=0$), in which the linear terms in the space variables cannot be eliminated by translation, due to the absence of quadratic terms.

\section[Maximal superintegrable class canonically conjugated to natural 2D systems]{Maximal superintegrable class canonically conjugated\\ to natural 2D systems}\label{sec:maximally_higher}

Let us consider the system whose magnetic field and effective potential read
\begin{gather}\label{magfieldB}
\vec B(\vec x)=(0, \gamma,0),\qquad \gamma\in\mathbb{R}\setminus\{0\}
\end{gather}
and
\begin{gather}\label{potB}
W(\vec x)= V(x_2),
\end{gather}
respectively. This system can be written in the form~\eqref{Sep1}, with the gauge chosen as
\begin{gather*}%\label{CaseBvecpot}
\vec A(\vec x)=(0,0, -\gamma x_1).
\end{gather*}
Its Hamiltonian reads
\begin{gather}\label{HB}
H=\frac12\big(p_1^2+p_2^2+p_3^2\big)-\gamma x_1 p_3 +\frac{\gamma^2}{2} x_1^2+ V(x_2).
\end{gather}
Actually by a different choice of the gauge and a canonical permutation of the variables $x_1$ and $x_2$ we see that the system belongs also to Case~II.
The Hamiltonian \eqref{HB} admits three independent first-order integrals~\cite{MS}
 \begin{gather}\label{integralsB}
 I_1 = p_1-\gamma x_3,\qquad I_2 = p_3, \qquad I_3 = 2l_2 +\gamma \big(x_1^2-x_3^2\big).
 \end{gather}
Out of them, we can construct two Cartesian-type integrals,
\begin{gather}\label{CartIntCan}
X_1=I_1^2+\gamma I_3,\qquad X_2=2 H-I_1^2-I_2^2 -\gamma X_3.
\end{gather}

The system can be reduced to two degrees of freedom through the following canonical transformation
\begin{gather}\label{reducingtrB}
 x_1= X+ \frac{P_3}{\gamma},\qquad x_2=Y,\qquad x_3=Z+\frac{1}{\gamma}P_1,\qquad p_j=P_j,\qquad j=1,2,3,
\end{gather}
with the second type generating function
\begin{gather*}
G(\vec x,\vec P)=\left(x_1-\frac {1}{\gamma} P_3\right) P_1+ x_2 P_2+ x_3 P_3.
\end{gather*}
The Hamiltonian in the new coordinates reads
\begin{gather}\label{HB2}
\mathcal K(\vec X,\vec P)=\frac12\big(P_1^2+P_2^2\big)+ \frac12 \gamma^2 X^2+ V(Y),
\end{gather}
i.e., it is effectively in two degrees of freedom and without magnetic field. This system \eqref{HB2} has two cyclic coordinates in the full phase space $(\vec X, \vec P)$, namely $Z$ and $P_3$, that are therefore both integrals. Expressed in the original variables, these integrals correspond to~$p_3$ and $\frac{I_1}{\gamma}$ as in~\eqref{integralsB}. Moreover~\eqref{HB2} separates in the Cartesian coordinates $(X,Y)$, and the corresponding Cartesian-type integrals, $\mathcal I_1$, $\mathcal I_2$, once written in the original coordinates, provide~\eqref{CartIntCan}. Thus we have
\begin{Proposition}\label{lemma:HBmaxsuper}
 The system with the magnetic field~\eqref{magfieldB} and potential~\eqref{potB} is maximally superintegrable if and only if~\eqref{HB2}, seen as a system in two degrees of freedom on the phase space $(X,Y,P_1,P_2)$ has one additional integral of motion, besides $\mathcal I_1$, $\mathcal I_2$, and independent of them.
\end{Proposition}
Therefore the problem of maximal superintegrability of \eqref{HB} has been reduced to the two-dimensional problem of superintegrability of \eqref{HB2}. In particular, all the potentials $V(Y)$ that make~\eqref{HB2} superintegrable give (by simply replacing $Y=x_2$) the effective potentials that render~\eqref{HB} superintegrable.

The cases
\begin{gather}\label{1/x2potential}
V(Y)=\frac{c}{Y^2}+\frac{\gamma^2 Y^2}{8},
\end{gather}
and
\begin{gather}\label{g22z2potential}
V(Y)=\frac{\gamma^2}{2} Y^2,
\end{gather}
that correspond to 3D superintegrable systems with additional second-order integral have already been found in~\cite{MS} with a different approach.

All the potentials $V(Y)$ that lead to second and third-order superintegrability in 2D have been classified~\cite{MiPoWin}.
If we focus on second-order integrals, they are listed in Table~\ref{table:2Dquad}.
The systems that can be obtained from it, after applying the transformation~\eqref{reducingtrB} and are still quadratically superintegrable are given by~\eqref{1/x2potential}, and
\begin{gather*}
V(Y)=\frac{\gamma^2}{2} Y^2 + c Y,
\end{gather*}
that, since $\gamma\neq0$, can be reduced to \eqref{g22z2potential} by translation in~$Y$.

However, higher-order superintegrable systems can be generated, e.g., from
\begin{gather}\label{B4thOrder}
V(Y)=\frac{c}{Y^2}+\frac{\gamma^2 Y^2}{2},\qquad c\geq 0.
\end{gather}
The additional integral of \eqref{HB2} is second order and reads (see Table \ref{table:2Dquad})
\begin{gather*}
\mathcal X_4= \mathcal L_3^2+ 2c\frac {X^2}{Y^2}.
\end{gather*}
Here $\mathcal L_3$ denotes the third component of the angular momentum with respect to the coordinates $(X,Y,Z, P_1,P_2,P_3)$.
Inverting the transformation \eqref{reducingtrB}, it gives the fourth-order integral
\begin{gather*}
 X_4=\frac{1}{\gamma^2}\left(\big(p_2 ^{A} p_3 ^{A}+\gamma p_1^{A} x_2\big)^2+2 c \frac{\big(p_3^{A}\big)^2}{x_2^{2}}\right).
\end{gather*}
Actually, by polynomial combinations with the other integrals, it can be reduced to the third order one
\begin{gather*}
 X_5 = 2 \gamma p_2^{A} p_3^{A} l_3^{A}+\gamma^2 \Big( x_1^2\big( p_2^{A}\big)^2 + x_2^2\Big( \big(p_3^{A}\big)^2-\big(p_1^{A}\big)^2 \Big)\Big)+ 2\gamma \frac{x_1}{x_2^2}\big(\gamma^2 x_2^4+ 2 c\big) p_3^A \\
\hphantom{X_5 =}{}+\gamma^2\frac{x_1^2}{x_2^2}\big(\gamma^2 x_2^4+2 c\big),
\end{gather*}
that cannot be further reduced to lower order by using any of the integrals~\eqref{integralsB} nor~\eqref{CartIntCan}.

A more general 3D infinite family of maximally superintegrable system, including the previous cases \eqref{1/x2potential} and~\eqref{B4thOrder} and the one found in \cite{MS2}, corresponds to the caged oscillator
\begin{gather}\label{cagedOsc}
V(Y)=\frac{c}{Y^2}+\frac{ m^2}{\ell^2} \gamma^2 Y^2,\qquad \ell,m\in\mathbb{N}
\end{gather}
If we compare it with \eqref{extcage2}, we see that for $\gamma^2=\omega l_2^2$, $\alpha_2=0$, $\beta_2=c$ and $m_2$ satisfying \eqref{ratio_cond} the two obtained 3D families would have the same scalar potential. However, the magnetic fields differ, rendering~\eqref{cagedOsc} maximally superintegrable, while~\eqref{extcage2}~-- as far as we can see~-- is only minimally superintegrable.

\section{Second-order integrals}\label{sec:second_ord_int}

Any second-order integral of motion we can write
\begin{gather}\label{classint}
 X= \sum_{j=1}^{3} h_j(\vec x) p_j^A p_j^A + \sum_{j,k,l=1}^{3} \frac{1}{2} |\epsilon_{jkl}| n_j(\vec x) p_k^A p_l^A + \sum_{j=1}^{3} s_j(\vec x) p_j^A+m(\vec x),
\end{gather}
where $\epsilon_{jkl}$ is the completely antisymmetric tensor with $\epsilon_{123}=1$.

The condition that the Poisson bracket
\begin{gather*}%\label{PoissonBracket}
\{a(\vec x,\vec p),b(\vec x,\vec p)\}=\sum_{j=1}^{3}\left(\frac{\partial a}{\partial {x_j}} \frac{\partial b}{\partial {p_j}} - \frac{\partial b}{\partial {x_j}} \frac{\partial a}{\partial {p_j}} \right)
\end{gather*}
of the integral \eqref{classint} with the Hamiltonian~\eqref{Hamiltonian} vanishes
\begin{gather*}
\{ H,X\}=0
\end{gather*}
seen as a polynomial in the momenta leads to the determining equations for the unknown functions $h_j$, $n_j$, $s_j$, $j=1,2,3$ and $m$ in the integral. Order by order (from the third to the zeroth) they read (cf.~\cite{MSW}):
\begin{gather}\label{3ordcond}
\begin{aligned}
&\partial_{x_1} h_1 = 0, \qquad && \partial_{x_2} h_1 = -\partial_{x_1} n_3 , \qquad && \partial_{x_3} h_1 =- \partial_{x_1} n_2 ,&\\
& \partial_{x_1} h_2 =-\partial_{x_2} n_3, \qquad && \partial_{x_2} h_2 =0, \qquad && \partial_{x_3} h_2 =-\partial_{x_2} n_1 ,& \\
& \partial_{x_1} h_3 =- \partial_{x_3} n_2 , \qquad && \partial_{x_2} h_3 =- \partial_{x_3} n_1 , \qquad && \partial_{x_3} h_3 = 0,\\
 & \nabla \cdot \vec n =0 ,&& && &
\end{aligned}\\
\nonumber \partial_{x_1} s_1 = n_2 B_2-n_3 B_3, \\
\nonumber \partial_{x_2} s_2 = n_3 B_3-n_1 B_1, \\
\nonumber \partial_{x_3} s_3 = n_1 B_1-n_2 B_2, \\
\label{2ordcond} \partial_{x_2} s_1 + \partial_{x_1} s_2 =n_1 B_2 -n_2 B_1+2 (h_1 - h_2) B_3, \\
\nonumber \partial_{x_3} s_1+\partial_{x_1} s_3 = n_3 B_1-n_1 B_3+2 (h_3 - h_1) B_2, \\
\nonumber \partial_{x_2} s_3+\partial_{x_3} s_2 = n_2 B_3-n_3 B_2+2 (h_2 - h_3) B_1,
\\
\nonumber \partial_{x_1} m = 2 h_1 \partial_{x_1} W+ n_3 \partial_{x_2} W+ n_2 \partial_{x_3} W+s_3 B_2-s_2 B_3, \\
\label{1ordcond} \partial_{x_2} m = n_3 \partial_{x_1} W+2 h_2 \partial_{x_2} W+ n_1 \partial_{x_3} W+s_1 B_3-s_3 B_1, \\
\nonumber \partial_{x_3} m = n_2 \partial_{x_1} W+ n_1 \partial_{x_2} W+2 h_3 \partial_{x_3} W+s_2 B_1-s_1 B_2,\\
\label{0ordcond}
\vec s \cdot \nabla W = 0.
\end{gather}
The equations \eqref{3ordcond} prescribe that the functions $h_j$, $n_j$ are such that the highest-order terms in the integral \eqref{classint} are linear combinations of products of the generators $p_1$, $p_2$, $p_3$, $l_1$, $l_2$, $l_3$ of the Euclidean group, where $l_j=\sum\limits_{k,l} \epsilon_{jkl} x_k p_l$~\cite{MSW}. Explicitly, in terms of the expressions~\eqref{covarP}, we have
\begin{gather}\label{classintUEA}
X =\sum_{i,j\colon i\leq j}\alpha_{ij}l_i^A l_j^A+ \sum_{i,j}\beta_{ij}p_i^A l_j^A+\sum_{i,j:\; i\leq j} \gamma_{ij}p_i^A p_j^A + \sum_{j=1}^{3} s_j(\vec x) p_j^A+m(\vec x),
\end{gather}
where $l_j^A=\sum\limits_{k,l} \epsilon_{jkl} x_k p_l^A$.
By subtracting the Hamiltonian and the two Cartesian integrals we can a priori set $\gamma_{11}=\gamma_{22}=\gamma_{33}=0$.
There are compatibility conditions on equations~\eqref{2ordcond}, consequence of the following conditions on the derivatives of the functions $s_j$, namely,
\begin{gather}
\partial^2_{x_2}\partial_{x_1}s_1+ \partial^2_{ x_1} \partial_{x_2}s_2=\partial_{x_1}\partial_{ x_2}(\partial_{x_2} s_1+ \partial_{x_1} s_2),\nonumber\\
\partial^2_{x_3}\partial_{x_1}s_1+ \partial^2_{ x_1} \partial_{x_3}s_3=\partial_{x_1}\partial_{x_3}(\partial_{x_3} s_1+ \partial_{x_1} s_3),\nonumber\\
\partial^2_{x_3} \partial_{x_2}s_2+ \partial^2_{ x_2} \partial_{x_3} s_3=\partial_{x_2}\partial_{x_3}(\partial_{x_3} s_2+ \partial_{x_2} s_3),\nonumber\\
\partial_{x_1}\partial_{x_3}(\partial_{x_2} s_1+\partial_{x_1} s_2)=2\partial_{x_2}\partial_{x_3}(\partial_{x_1}s_1)-\partial_{x_1}\partial_{x_2}(\partial_{x_3} s_1+\partial_{x_1} s_3)+\partial^2_{x_1}(\partial_{x_3} s_2+\partial_{x_2} s_3),\nonumber\\
\partial_{x_2}\partial_{x_3}(\partial_{x_2} s_1+\partial_{x_1} s_2)=2\partial_{x_1}\partial_{x_3}(\partial_{x_2}s_2)-\partial_{x_1}\partial_{x_2}(\partial_{x_3} s_2+\partial_{x_2} s_3)+\partial^2_{x_2}(\partial_{x_3} s_1+\partial_{x_1} s_3),\nonumber\\
\partial_{x_2}\partial_{x_3}(\partial_{x_3} s_1+\partial_{x_1} s_3)=2\partial_{x_1}\partial_{x_2}(\partial_{x_3}s_3)-\partial_{x_1}\partial_{x_3}(\partial_{x_3} s_2+\partial_{x_2} s_3)+\partial^2_{x_3}(\partial_{x_2} s_1+\partial_{x_1} s_2).\!\!\!\!\label{comps}
\end{gather}
These translate into compatibility conditions on the magnetic field and the constants in the coefficients of the second-order terms.
Further compatibility constraints come from \eqref{1ordcond}, consequence of
\begin{gather}\label{compm}
\partial_{x_i}\partial_{x_j} m=\partial_{x_j}\partial_{x_i} m,\qquad i,j=1,2,3, \quad i\neq j.
\end{gather}

\section{A necessary condition for second-order superintegrability}\label{sec:necessary}
Both classes of systems that separate in Cartesian coordinates have at least one first-order integral and it is always possible to choose a gauge so that such integral reads as one of the linear momenta. To fix the ideas, let us work in such a gauge choice and assume that the constant momentum is $p_3$. If a second-order integral $X$ exists, then $K_1=\{X,p_3\}$ is still an integral at most of second order or a constant. Since the highest-order terms in $X$ are as in~\eqref{classintUEA}, they can be at most quadratic in $x_3$. This means that if $K_1$ is quadratic in the momenta, its second-order terms are at most linear in $x_3$, since $K_1=\{X,p_3\}=\frac{\partial X}{\partial x_3}$. Thus, $K_2=\{K_1,p_3\}$ can be, as above, either an integral at most quadratic or a constant. If $K_2$ is again quadratic, $K_3=\{K_2,p_3\}$ can be now at most linear in the momenta, since the highest-order terms in $K_2$ do not depend on $x_3$. Therefore, we can conclude that if a second-order independent integral $X$ exists, then necessarily there must exist a second-order integral (which could be $X$ itself) such that $\{X,p_3\}$ is at most linear in the momenta. In general, for a conserved momentum~$p_j$, the result is the same, it is enough to replace $x_3$ by $x_j$ in the argument above. Thus, we obtain the following
\begin{Proposition}\label{lemma:linear}
Let the system defined by $H$ as in \eqref{Hamiltonian} separate in Cartesian coordinates and have a quadratic integral $\mathcal I$ independent of the Cartesian integrals. Then there exists a second-order integral~$X$, not necessarily different from $\mathcal I$, such that $\{X,p_j\}$ is a polynomial expression in the momenta of at most first order, for some~$j$.
\end{Proposition}
Thus, to answer the question on the existence of an additional second-order integral for the class of systems we are considering here, we can start by answering the simpler question on the existence of the necessary integral $X$ that satisfies the above property. This is done in the following Sections~\ref{sec:case1} and~\ref{sec:case2} and Appendices~\ref{appendix:case1},~\ref{appendix:case1-part2},~\ref{appendix:case2}.

Since we found that the special case in which the magnetic field is constant and the functions~$V_j$ are second-order polynomials in the respective variables appears several times in the computation therein, we discuss it at once in the separate Section~\ref{sec:pol2}.

\section{Quadratic superintegrability in Case I}\label{sec:case1}
We start with the class of systems in~\eqref{Omega_sep1}. To fix the ideas, let us choose a gauge as in~\eqref{Sep1} and assume that there exists a quadratic independent integral~$\mathcal I$. Thus, by Proposition~\ref{lemma:linear} there exist another quadratic integral~$X$ such that $\{X,p_3\}$ is at most first order as a polynomial in the momenta. Here we consider only the case in which the two Cartesian-type integrals do not reduce to first-order integrals. In case one of them does, then the system is at the intersection of Case~I and Case~II (up to a permutation of indices) and it is treated at once in Section~\ref{sec:case2}. Moreover, we assume there does not exist a linear integral, other than~$p_3$. If it exists, the corresponding systems can be found in~\cite{MS}, where there is a complete study of quadratically superintegrable systems with Cartesian integrals and one independent first-order integral.
We can have several cases:
\begin{itemize}\itemsep=0pt
\item [(i)] $\{X,p_3\}$ is at most linear and not vanishing. Thus the only possibility of finding something new is in assuming that $\{X,p_3\}$ is a dependent integral or a constant (we excluded the case there is an independent first-order integral). We therefore look for a quadratic integral $X$ such that
\begin{gather}\label{linearint0}
\partial_{x_3} X=\{X,p_3\}=c_1 p_3+ c_0,\qquad c_j\in\mathbb{R},
\end{gather}
and $c_j$ not both vanishing, $j=0,1$.
\item [(ii)] $\{X,p_3\}=0$ and there exist no quadratic integral independent of the Cartesian integrals and commuting with~$p_3$. Then~$X$ is trivial, in the sense that it depends on the Cartesian integrals and~$p_3$. However, to have a quadratic superintegrable system, a quadratic integral~$\mathcal I$ as in Proposition~\ref{lemma:linear} must exist. Without loss of generality, we can assume $X=\{ \mathcal I, p_3\}$ with
\begin{gather}\label{quadint00}
\{\mathcal I, p_3\}=a_0 p_3^2 + a_1 X_1+ a_2 X_2 + c_1 p_3 + c_0,
\end{gather}
where $X_1$ and $X_2$ are as in~\eqref{CartInt1}, $a_0,a_1,a_2,c_0,c_1\in\mathbb{R}$, not all $a_j$ vanishing, otherwise we are in the previous point i).
\item [(iii)] $\{ X,p_3\}=0$ and $X$ is independent of the Cartesian integrals. Since $X$ commutes with $p_3$, it satisfies the assumptions of Proposition~\ref{lemma_minimally}. Thus, the corresponding systems can be found in Table~\ref{table:2Dquad}.
If an additional quadratic independent integral exists, then its Poisson bracket with~$p_3$ cannot vanish. This is a consequence of the fact that the 2D system~\eqref{2dof} cannot have more than~$3$ independent integrals.
However, as in the previous point, there could exist a quadratic independent integral $\mathcal I$ such that $\{ \mathcal I, p_3\}$ depends on the others, namely
\begin{gather}\label{quadint0}
\{\mathcal I, p_3\}=a_0 p_3^2 + a_1 X_1+ a_2 X_2 + a_3 X_3+ c_1 p_3 + c_0,
\end{gather}
where $a_0,a_1,a_2,a_3,c_0,c_1\in\mathbb{R}$ and not all~$a_j$ are vanishing (otherwise we are in case~(i)), $X_1$, $X_2$ as in~\eqref{CartInt1} and $X_3=X$.
\end{itemize}

Let us investigate the possibilities for $X_3$ in \eqref{quadint0}. Its highest-order terms should come from a Poisson bracket of the quadratic terms of~$\mathcal I$ with~$p_3$, i.e., their derivatives with respect to~$x_3$. Moreover, by assumption~$X_3$ does not depend on~$x_3$. Thus, its second-order terms can arise only by taking derivatives of a second-order polynomial that contains terms of the form $p_i\cdot l_j$, $i=1,2,3$, $j=1,2$. By computing their Poisson bracket with $p_3$, we see that the only outcome (for an integral~$X_3$ independent of~$X_1$ and~$X_2$) is in terms of the type~$p_i p_{\ell}$, $i\neq\ell$. Looking at the integrals of the of 2D systems in Table~\ref{table:2Dquad}, and the dependent integrals obtained by their Poisson bracket with the Cartesian integrals, we see that the only possibility is~\eqref{quad_maximally} below.

Now that we outlined all the possibilities, we need to solve the determining equations \eqref{2ordcond}--\eqref{0ordcond}, for the different cases. For this, it is necessary to work in gauge covariant setting.

The conditions \eqref{linearint0}, \eqref{quadint00} and \eqref{quadint0} can be written together as (we can now set $a_3=0$):
\begin{gather}\label{linearint}
\partial_{x_3} X=a_0 \big(p_3^A-u_1(x_2)+ u_2 (x_1)\big) ^2 + a_1 X_1+ a_2 X_2+ c_1 \big(p_3^A-u_1(x_2)+ u_2 (x_1)\big) + c_0,\!\!\!
\end{gather}
where with an abuse in the notation we denoted~$\mathcal I$ as $X$ (the unknown independent integral we are looking for), with $a_j,c_j\in\mathbb{R}$ and not all vanishing. For $a_j=0$, $j=0,1,2$ we are in case~(i).

Equation \eqref{linearint} implies the following values for the second-order terms of $X$ as in \eqref{classintUEA}:
\begin{gather}
\alpha_{11}=\alpha_{22}=\alpha_{12}=\alpha_{13}=\alpha_{23}=\beta_{31}=\beta_{32}=0, \qquad \beta_{11}=\beta_{22},\nonumber\\
 a_0=0,\qquad a_1=\beta_{12},\qquad a_2=-\beta_{21}.\label{coeffcond1}
\end{gather}
Moreover, since $\vec p\cdot \vec L=0$ we can set $\beta_{22}=0$ (and consequently also $\beta_{11}=0$).

Concerning the lower-order terms, by integrating the right-hand side of~\eqref{linearint}, we obtain the following restriction on the structure of~$X$:
\begin{gather}
s_j =S_j(x_1,x_2),\qquad j=1,2,\nonumber\\
s_3 =S_3(x_1,x_2)- (2\beta_{12} u_2(x_2)+2\beta_{21}u_1(x_2)-c_1) x_3,\nonumber\\
m_3 =c_0 x_3 + (u_1(x_2)-u_2(x_1)) \left((2\beta_{12}u_2(x_1)+2\beta_{21}u_1(x_2)- c_1) x_3\right)\nonumber\\
\hphantom{m_3 =}{} +(2\beta_{12} V_1(x_1)- 2\beta_{21} V_2(x_2))x_3 + M(x_1,x_2).\label{CaseIsm}
\end{gather}
With this simplifications at hand, we can solve equations \eqref{2ordcond}--\eqref{0ordcond}.

Let as assume that $a_1$ and $a_2$ in \eqref{linearint} are not both zero; e.g., let it be $a_1\neq0$. Then we can shift both the potential $V_1(x_1)$ and the third component of the vector potential by a constant, thus absorbing the constants~$c_0$ and~$c_1$.
Similarly, if $a_2\neq0$ we could use~$X_2$.
Therefore, by~\eqref{coeffcond1}, we see that if either $\beta_{12}\neq0$ or $\beta_{21}\neq0$, we can proceed in the solution of \eqref{2ordcond}--\eqref{0ordcond} as if $c_1=c_0=0$. We obtain that no new superintegrable system can be found in this case.
 The details of the computation are in Appendix~\ref{appendix:case1-part2}.

For $\beta_{12}=\beta_{21}=0$ we find it convenient to start from~\eqref{2ordcond}, in which the third equation simplifies to
\begin{gather}
(\beta_{33} x_1 + \gamma_{23})u_1'(x_2)+ (\beta_{33} x_2 - \gamma_{13}) u_2'(x_1)-c_1 =0.\label{eqtostart}
\end{gather}
The above equation could be trivially satisfied for some of the functions $u_j$ or not. This determines a major splitting in the computation. For the details see Appendix~\ref{appendix:case1}, the resulting list of systems is given in the conclusions, Section~\ref{Conclusion1}.

\section{Quadratic superintegrability in Case II}\label{sec:case2}
For the class of systems \eqref{Omega_sep2} we can choose a gauge so that there are two mutually orthogonal conserved linear momenta. Let us assume that they are $p_2$ and $p_3$ as in~\eqref{CartInt2}. As above, we assume there exists an independent quadratic integral. Thus, by Proposition~\ref{lemma:linear} we can
have two possibilities:
\begin{itemize}\itemsep=0pt
\item [(i)] there exists a quadratic integral $X$ such that $\{X,p_2\}=\{X,p_3\}=0$. Then $X$ is an integral of the reduced system obtained from~\eqref{Omega_sep2} by setting the conserved momenta to constants, i.e., function of the 1-dimensional Hamiltonian. Thus, it is dependent on the Hamiltonian and the conserved momenta. The only hope to find something interesting is to look for a~quadratic integral $\mathcal I$ such that $ \{\mathcal I, p_j\}=X$ for some~$j$.
\item [(ii)] There exists a quadratic integral $X$ such that $\{X,p_j\}$ is linear and not vanishing for at least one $p_j$, $j=2,3$. Without loss of generality we can assume that $\{X,p_3\}\neq 0$, otherwise we permute the coordinates~$x_2$ and~$x_3$.
\end{itemize}
Let us set $j=3$ in both cases and with an abuse of notation let us rename~$\mathcal I$ in case~(i) as~$X$. Thus, we look for a quadratic integral~$X$ such that{\samepage
\begin{gather}\label{Comm2}
\{X, p_3\}= 2 a_0 \big(H-X_1^2-X_2^2\big)+a_1 X_1^2+a_2 X_2^2+a_3 X_1 X_2 + c_0 + c_1 X_1+ c_2 X_2,
\end{gather}
$X_1$, $X_2$ as in~\eqref{CartInt2}. For $a_j=0$, $j=1,\dots,4$, we have case~(ii).}

Equation \eqref{Comm2} implies the following conditions on the coefficients of the higher-order terms of the integral, expressed as in \eqref{classintUEA} (again, we use the condition $\vec p \cdot \vec L=0$)
\begin{gather}
\alpha_{11} =\alpha_{22}=\alpha_{12}=\alpha_{13}=\alpha_{23}=\beta_{11}=\beta_{22}=\beta_{32}=0,\nonumber\\
a_0 = \beta_{12},\qquad a_1=-\beta_{21},\qquad a_2=0,\qquad a_3=-\beta_{31}.\label{alphaComm2}
\end{gather}
Moreover, by subtracting $X_1X_2$ from $X$, we can set $\gamma_{23}=0$.

Still as a consequence of \eqref{Comm2}, we have further conditions on the coefficients of the lower-order terms
\begin{gather*}
s_1=S_1(x_1,x_2),\qquad s_2=S_2(x_1,x_2)+ (2(\beta_{12}+ \beta_{21}) u_3(x_1)-\beta_{31} u_2(x_1)+ c_1) x_3,\nonumber\\
s_3=S_3(x_1,x_2)+ z( \beta_{31} u_3(x_1)-2 \beta_{12}u_2(x_1)+c_2 )%\label{ssolComm2}
\end{gather*}
and
\begin{gather*}
m =M(x_1,x_2) -\big((2\beta_{12}+\beta_{21})u_3(x_1)^2- \beta_{31}u_2(x_1)u_3(x_1) \nonumber\\
\hphantom{m =}{}+ c_1 u_3(x_1)- c_2 u_2(x_1)+2\beta_{12}u_2(x_1)^2 -2\beta_{12} V_1(x_1)- c_0\big)x_3.%\label{msolComm2}
\end{gather*}
With these simplifications at hand, we are able to solve the determining
equations~\eqref{3ordcond}--\eqref{0ordcond}.

Let us perform the substitution
\begin{gather}\label{USub}
u_j(x_1)=U_j'(x_1),\qquad j=2,3.
\end{gather}
Since $u_j$ are defined in~\eqref{Omega_sep1} up to addition of arbitrary constants and~$U_j$ is defined as in~\eqref{USub}, in the following we can set to zero all the coefficient of first and zero-order powers of~$x_1$ in the solutions for~$U_j$.

From \eqref{2ordcond} we find
\begin{gather*}
S_1(x_1,x_2) =s_1(x_2)+\beta_{12}U_2(x_1)+ (\beta_{13}-2\alpha_{33} x_2)U_3(x_1) -(\beta_{12} x_1 +\beta_{33}x_2-\gamma_{13})U_2'(x_1)\nonumber\\
\hphantom{S_1(x_1,x_2) =}{} + (2\alpha_{33} x_1 x_2-\beta_{13}x_1+\beta_{23} x_2- \gamma_{12})U_3'(x_1),\nonumber\\
S_2(x_1,x_2)= s_2(x_1)- \left(\alpha_{33}x_1 x_2^2-\beta_{13}x_1 x_2+\frac{1}{2}\beta_{23} x_2^2-\gamma_{12} x_2 \right)U_3''(x_1),\nonumber\\
S_3(x_1,x_2)=s_3(x_1)+ c_1 x_2+\beta_{31}x_2 U_2'(x_1) -2 (\beta_{12}+\beta_{21})x_2 U_3'(x_1)\nonumber\\
\hphantom{S_3(x_1,x_2)=}{} + \left(\alpha_{33} x_1 x_2^2-\beta_{13} x_1 x_2 +\frac12 \beta_{23} x_2^2-\gamma_{12} x_2 \right)U_2''(x_1)\nonumber\\
\hphantom{S_3(x_1,x_2)=}{}- \left(\beta_{12} x_1 x_2 +\frac12 \beta_{33}x_2^2-\gamma_{13} x_2\right)U_3''(x_1),
\end{gather*}
where $U_j$ and $s_{\ell}$ must satisfy the third, fourth and fifth equation of~\eqref{2ordcond}.
Let us continue by considering the third of these equations, namely
\begin{gather}
 U_2''(x_1) (\beta_{12}x_1+\beta_{33} x_2-\gamma_{13})+2 \beta_{12} U_2'(x_1)-\beta_{31}U_3'(x_1)-c_2=0 \label{eqtostart2},
\end{gather}
together with the compatibility conditions~\eqref{comps}. The first one is trivially satisfied, while the remaining five read
\begin{gather}
\beta_{33} U_2'''(x_1)=0,\nonumber\\
(2\alpha_{33} x_1+\beta_{23}) U_2'''(x_1)+6\alpha_{33}U_2''(x_1)-\beta_{33} U_3'''(x_1)=0,\nonumber\\
(\beta_{12}x_1+\beta_{33}x_2- \gamma_{13}) U_2^{(4)}(x_1)+4 \beta_{12} U_2'''(x_1)-\beta_{31} U_3'''(x_1)=0,\nonumber\\
-(2\alpha_{33} x_1 x_2 -\beta_{13} x_1 +\beta_{23} x_2- \gamma_{12})U_3^{(4)}(x_1)+
4(\beta_{13}-2 \alpha_{33} x_2) U_3'''(x_1)-\beta_{21} U_2'''(x_1)=0,\nonumber\\
 (8 \alpha_{33}x_2-4\beta_{13}-\beta_{31})U_2'''(x_1)- (4\beta_{12} +\beta_{21}) U_3'''(x_1) \label{COmegaU2}\\
\qquad{} +(2 \alpha_{33} x_1 x_2-\beta_{13}x_1+\beta_{23}x_2- \gamma_{12}) U_2^{(4)}(x_1)-
(\beta_{12}x_1 +\beta_{33}x_2-\gamma_{13}) U_3^{(4)}(x_1)=0.\nonumber
\end{gather}
We can have different subcases according to whether the equations \eqref{eqtostart2},~\eqref{COmegaU2} are trivially sa\-tisfied for some of the functions $U_j$ or not. This determines a major splitting in the computation. The details are given in Appendix \ref{appendix:case2}.

\section[Constant magnetic field and second-order polynomial potentials]{Constant magnetic field\\ and second-order polynomial potentials}\label{sec:pol2}

Let us consider the particular case in which the magnetic field is constant
\begin{gather}\label{Bpol2}
\vec B(\vec x)=(a_1,a_2,0)
\end{gather}
 and in \eqref{Omega_sep1} we have
\begin{gather*}
V_1(x)= v_{11} x_1 + v_{12} x_1^2,\qquad V_2(x_2)=v_{21}x_2 + v_{22} x_2^2,\qquad u_1= a_1 x_2,\qquad u_2= -a_2 x_1.
\end{gather*}
This system appears in various branches of calculation in the appendices; thus we find it practical to discuss it separately here.

Notice that since the magnetic field is constant, by rotation around $x_3$-axis we could reduce it to the case in which it is aligned with one of the Cartesian axis. However, the system would no longer separate in the corresponding rotated Cartesian coordinates, therefore we prefer not to perform such a rotation.

Let us also point out that if $V_1(x_1)=0$, for constant magnetic field a rotation around $x_2$ brings the system~\eqref{Omega_sep1} into~\eqref{Omega_sep2}. Thus, what we will deduce in the following for $V_1=0$ applies also for~\eqref{Omega_sep2}.

\subsection[$v_{12}$ and $v_{22}$ both not vanishing]{$\boldsymbol{v_{12}}$ and $\boldsymbol{v_{22}}$ both not vanishing}

Let us assume $v_{12}$ and $v_{22}$ are both not vanishing. Then by the translation of the coordinate system we can set $v_{11}=v_{21}=0$ without loss of generality.

Then, similarly to Section~\ref{sec:maximally_higher}, we can reduce to a natural Hamiltonian system through canonical transformations. Namely, let us take as the generating function
\begin{gather}\label{v12v22cantransf}
G= \left(x_1-\frac{a_2 P_3}{2 v_{12}}\right)P_1+\left(\frac{a_1 P_3}{2v_{22}}+x_2\right)P_2+x_3 P_3,
\end{gather}
so that $p_j=P_j$, $j=1,2,3$ and
\begin{gather*}
x_1= X+\frac{a_2 P_3 }{2 v_{12}},\qquad x_2=Y-\frac{a_1 P_3}{2v_{22}}, \qquad x_3= Z+\frac{a_2 v_{22} P_1-a_1 v_{12} P_2}{2 v_{12}v_{22}}.
\end{gather*}
After the transformation, with gauge chosen as in~\eqref{Sep1}, the Hamiltonian reads
\begin{gather*} %\label{oscillator}
H=\frac12\left(P_1^2+P_2^2+ \left(1-\frac{a_1^2}{2v_{22}}-\frac{a_2^2}{2v_{12}}\right)P_3^2\right)+ v_{12} X^2 +v_{22} Y^2.
\end{gather*}
If $\frac{a_1^2}{2v_{22}}+\frac{a_2^2}{2v_{12}}\neq1$ we can, by a canonical transformation
\begin{gather*}
P_3= \frac{1}{\lambda} \tilde{P}_3, \qquad Z= \lambda \tilde{Z}, \qquad \lambda^2= \left| 1-\frac{a_1^2}{2v_{22}}-\frac{a_2^2}{2v_{12}} \right|
\end{gather*}
scale the $P_3^2$ term to have the Hamiltonian of the form
\begin{gather*} %\label{oscillatorscaled}
H=\frac12\big(P_1^2+P_2^2 \pm \tilde{P}_3^2\big)+ v_{12} X^2 +v_{22} Y^2.
\end{gather*}
The system can therefore be reduced to a system determined by a two-dimensional, possibly inverted, anisotropic harmonic oscillator and free motion along the $Z$-direction. The original 3D system is minimally superintegrable if and only if the corresponding 2D oscillator is superintegrable as a system in the $(X,Y,P_1,P_2)$ space. If $v_{12}=v_{22}$ we have a special case of the system~$\mathcal{E}_3$ in Table~\ref{table:2Dquad}. If $\frac{v_{12}}{v_{22}} \in \mathbb{Q}$, $\frac{v_{12}}{v_{22}}\neq 1$ we have a higher-order integral when expressed in the variables $x_j$, $p_j$.

If $\frac{a_1^2}{2v_{22}}+\frac{a_2^2}{2v_{12}}=1$, the coordinate $P_3$ becomes cyclic and its conjugated variable $Z$ is an independent constant of motion. In this case the system \eqref{Omega_sep1} becomes at least minimally superintegrable. It is maximally superintegrable if and only if its reduction on the $(X,Y,P_1,P_2)$ space is superintegrable. Indeed, we have reduced to the system \eqref{HB2} for $V(Y)=v_{22} Y^2$ and $2v_{12}=\gamma^2=a_2^2$. Its maximally superintegrable exception is included in the family of systems~\eqref{cagedOsc}.

\subsection[$v_{22}$=0 and $v_{12}$ not vanishing]{$\boldsymbol{v_{22}=0}$ and $\boldsymbol{v_{12}}$ not vanishing}

In this case by translation in $x_1$ we can still set $v_{11}=0$. Then by a canonical transformation such that
\begin{gather*}
x_1=X + \frac{a_2}{2 v_{12} } P_3,\qquad x_2=Y,\qquad x_3=Z+ \frac{a_2}{2 v_{12} }P_1,\qquad p_j=P_j,\qquad j=1,2,3,
\end{gather*}
we obtain the system
\begin{gather*}
H=\frac12 \big(P_1^2 + P_2^2 + P_3^2\big) + v_{12} X^2 + a_1 P_3 Y + v_{21} Y.
\end{gather*}
If $a_1=0$ we have reduced to a natural system. By reducing the integral~$P_3$ we have a~2D system that, to our knowledge, is not superintegrable.

If $a_1\neq0$ we have reduced to the case with magnetic field aligned along one axis. The effective potential of the so obtained system reads
\begin{gather}\label{Wpol21}
W=v_{12} X^2 + v_{21} Y - \big(a_1^2 Y^2\big)/2.
\end{gather}
Thus, by the translation
\begin{gather*}
Y\rightarrow Y+ \frac{v_{21}}{a_1^2}
\end{gather*}
 we can eliminate the linear term from the effective potential. A~shift of the vector potential by a constant, i.e.,
\begin{gather*}
 A_3\rightarrow A_3-\frac{v_{21}}{a_1},
\end{gather*}
gives
\begin{gather*}
H=\frac12 \big(P_1^2 + P_2^2 + P_3^2\big) + v_{12} X^2 + a_1 P_3 Y.
\end{gather*}
Thus, without loss of generality, we can set $v_{21}=0$. By plugging~\eqref{Wpol21} and~\eqref{Bpol2} with these simplifications into the determining equations for a second-order integral as in~\eqref{classintUEA}, we find that they have no solution.

\subsection[$v_{12}=v_{22}=0$]{$\boldsymbol{v_{12}=v_{22}=0}$}

We have a subcase of the system $\mathcal{E}_3$ in Table~\ref{table:2Dquad}.
Thus, the system admits a second-order integral. With the gauge chosen as in~\eqref{Sep1}, that integral reads
\begin{gather*}%\label{int_quad_1}
X_3=p_1 p_2+ (v_{11}- a_2 p_3)x_2+ (v_{21}+a_1 p_3) x_1.
\end{gather*}
Equivalently, in gauge covariant form, we have
\begin{gather*}
X_3=p_1^A p_2^A+ (a_1 x_1-a_2 x_2) p_3^A- \big(a_1^2+a_2^2\big)x_1 x_2+a_1 a_2\big(x_1^2+x_2^2\big)+v_{11}x_2+v_{21}x_1,
\end{gather*}
corresponding to the fact that the system actually separates in any rotated system of Cartesian coordinates, since the Hamiltonian is linear in the space variables. Without altering the structure of the Cartesian-type integrals, we can therefore by rotation align the magnetic field along one Cartesian axis, let us say the $x_2$-axis. Thus, without lost of generality, let us assume $a_1=0$. The determining equations for an additional second-order integral can be solved. We find for $v_{11}=0$ one maximally superintegrable system:
\begin{gather}\label{quad_maximally}
\vec B(\vec x)=(0,a_2,0),\qquad W(\vec x)=v_{21} x_2-\frac12 a_2^2 x_1^2,
\end{gather}
with the integral
\begin{gather*}%\label{int_quad_2}
X_4 =3p_3^A l_1 ^A-p_1^A l_3^A-\frac{3v_{21}}{a_2} l_2^A + a_2 x_1 x_2 p_3^A + 3 a_2 x_1 l_1^A + v_{21} x_1^2 + a_2^2 x_1^2 x_2\nonumber\\
\hphantom{X_4}{} =3p_3 l_1- p_1 l_3-\frac{3}{a_2}\big(3 v_{21} l_2+2 a_2^2 x_1 x_2 p_3 +2 a_2 v_{21} x_1^2\big).
\end{gather*}

\section{Conclusions}\label{Conclusion}

Let us summarize our results. We have provided an exhaustive determination of quadratically superintegrable systems which separate in the Cartesian coordinates with magnetic field. In addition, we have found classes of systems minimally and maximally superintegrable with higher-order integrals. We list them below for reader's convenience.

\subsection{Superintegrable systems with second-order integrals}\label{Conclusion1}
We have constructed an exhaustive list of quadratically superintegrable systems with nonvanishing magnetic field which separate in Cartesian coordinates. Under the assumption that there is no independent first-order integral other than the Cartesian ones (in that case we refer the reader to our previous work~\cite{MS,MSW}) we have found~8 classes of minimally superintegrable systems, among which one contains a quadratically maximally superintegrable subclass, cf.~\eqref{quad_maximally}.

For brevity, we write here the magnetic field, the electrostatic potential and the leading order terms in the integral(s) together with the reference to the equation in which the system was introduced. We refer the reader to the relevant formulas therein encoding the complete information about the integral(s).

\begin{description}
\item[Case I,] i.e., the magnetic field and potential are of the form~\eqref{Omega_sep1} and the Cartesian integrals as in~\eqref{CartInt1}. The superintegrable systems read
\begin{description}
\item[(a)] \begin{gather*}
\vec B(\vec x) = \big( a {\mathrm{e}}^{b x_2}, c, 0\big),\qquad
 W(\vec x)= a \left( w +\frac{ c}{b} x_1 \right){\mathrm{e}}^{b x_2} -\frac{a^2}{2 b^2}{\mathrm{e}}^{2 b x_2},
\end{gather*}
$ X_3 = p_1^A p_3^A + \cdots$, cf.~\eqref{BW31}.
\item[(b)]
\begin{gather*}
\vec B(\vec x)   = 2 \left(a_1 x_2-\frac{a_3}{x_2^3}, -a_1 x_1+ \frac{a_2}{x_1^3}, 0 \right),\nonumber \\ \nonumber
W(\vec x) = -\frac{1}{2} a_1^2 \big(x_1^2+x_2^2\big)^2-\frac{a_2^2}{2 x_1^4}-\frac{a_3^2}{2 x_2^4}-
 a_1 \left( a_2 \frac{x_2^2}{x_1^2}+ a_3\frac{x_1^2}{x_2^2}\right)  \\
  \hphantom{W(\vec x) =}{} -\frac{a_2 a_3}{x_2^2 x_1^2}+\frac{b_3}{x_2^2}+b_1 \big(x_1^2+x_2^2\big) +\frac{b_2}{x_1^2},
\end{gather*}
$X_3=\big(l_3^A\big)^2+\cdots$, cf. Table~\ref{table:2Dquad}.
\item[(c)] \begin{gather*}
\vec B(\vec x)   = \left( 2 \left( a_1 x_2- \frac{a_3}{x_2^3} \right),-8 a_1 x_1-a_2,0\right), \nonumber \\
W(\vec x)   = -\frac{a_1^2}{2} \big(4 x_1^2+x_2^2\big)^2 -\frac{a_2^2}{2} x_1^2 -\frac{a_3^2}{2 x_2^4} -a_2 a_3 \frac{x_1}{x_2^2}   \\
\hphantom{W(\vec x)   =}{} - a_1 a_2 x_1 \big(4 x_1^2+x_2^2\big)-4 a_1 a_3\frac{x_1^2}{x_2^2} + \frac{b_3}{x_2^2}+ b_1 \big(4 x_1^2+x_2^2\big)+b_2 x_1,
\end{gather*}
$X_3= p_2^A l_3^A +\cdots$, cf.\ Table~\ref{table:2Dquad}.
\item[(d)]
\begin{gather*}
\vec B(\vec x)   =  ( 2 a_1 x_2+a_3, -2 a_1 x_1-a_2,0  ), \\
W(\vec x)   = -\frac{a_1^2}{2} \left(x_1^2+x_2^2\right)^2-\frac{a_3^2}{2} x_2^2-\frac{a_2^2}{2} x_1^2 - a_2 a_1 x_1 \left(x_1^2+x_2^2\right) \\
\hphantom{W(\vec x)   =}{} -a_2 a_3 x_1 x_2-a_1 a_3 x_2 \big(x_1^2+x_2^2\big) + b_1 \big(x_1^2+x_2^2\big)+b_2 x_1+b_3 x_2,
\end{gather*}
$X_3=p_1^A p_2^A+\cdots$, cf.\ Table~\ref{table:2Dquad}.
When $a_1=a_3=0$ and $b_1=b_2=0$ the system becomes maximally superintegrable, with the additional integral of the form $X_4= p_1^A l_3^A-3p_3^A l_1 ^A+\cdots$, cf.~\eqref{quad_maximally}.
\end{description}
\item[Case II,] i.e., the magnetic field and potential are of the form~\eqref{Omega_sep2} and the Cartesian integrals as in~\eqref{CartInt2}. The superintegrable systems read
\begin{description}
\item[(a)] \begin{gather*}
\vec B(\vec x)=\big(0,a {\rm e}^{b x_1},0\big),\qquad W(\vec x)= w x_1+ c e ^{b x_1}-\frac12 \frac{a^2}{b^2} {\rm e}^{2 bx_1},
\end{gather*}
$X_3= p_1^A p_2^A - b p_3^A l_1^A +\cdots$, cf.~\eqref{BWSCa},
\item[(b)] \begin{gather*}
\vec B(\vec x)=\big(0,a (b-2) x_1^{b-3},0\big),\qquad W(\vec x)=-\frac{a^2 x_1^{2(b-2)}}{2}+ a (b-2) c x_1^{b-2}+ \frac{w}{x_1^2},
\end{gather*}
$X_3= p_1^A l_3^A - b p_3^A l_1^A +\cdots$, cf.~\eqref{BWSCb},
\item[(c)] \begin{gather*}
\vec B(\vec x)=\left(0,\frac{a}{x_1},0\right), \qquad W(\vec x)=-\frac12 a^2 \left(\ln |x_1|\right)^2 + b \ln |x_1|+ \frac{w}{x_1^2},
\end{gather*}
$X_3= 2 p_1^A l_3^A - p_3^A l_1^A +\cdots$, cf.~\eqref{BWSCc},
\item[(d)] \begin{gather*}
\vec B(\vec x) = \left( 0, 0, \frac{a}{x_1^3} \right), \qquad W(\vec x) = -\frac{a b \ln|x_1|}{x_1^2}-\frac{a^2}{8 x_1^4}+\frac{w}{x_1^2},
\end{gather*}
$X_3= p_1^A l_2^A +\cdots$, cf.~\eqref{C42UBW}.
\end{description}
\end{description}

Our approach also demonstrates that any quadratically maximally superintegrable system with magnetic field which separates in Cartesian coordinates would necessarily appear at the intersection of the presented classes. Given the different structure of the magnetic field in each of the cases we find only few potential candidates. One is the intersection of Case~I.d and Case~II.b which, as we already observed, leads to the maximally superintegrable system~\eqref{quad_maximally}. Another is the system Case I.a which for $c=0$ reduces to the system Case~II.a (upon interchange of the~$x_1$ and~$x_2$ coordinates and momenta). However, the integral $X_3$ of Case~I.a when $c=0$ becomes a function of the two first-order Cartesian integrals, i.e., it is not independent anymore. Last but not least, the systems Case~I.b, Case~I.c, Case~II.b and Case~II.d (after a permutation of coordinates) overlap for $a_1=a_2=0$ (Case I.b/c) and $b=0$ (Case~II.b/d) but the integrals again turn our to be dependent.

Thus we conclude that no other quadratically maximally superintegrable systems separating in Cartesian coordinates other than~\eqref{quad_maximally} and the ones found in~\cite{MS} exist.

\subsection{Superintegrable systems with higher-order integrals}\label{Conclusion2}

Above we have provided a complete answer to the problem of quadratic superintegrability for the considered classes of systems~\eqref{Omega_sep1} and~\eqref{Omega_sep2}. As we have seen, maximal superintegrability via at most quadratic integrals is very rare in the presence of magnetic field, as opposed to numerous purely scalar maximally superintegrable systems discussed, e.g., in~\cite{Evans1, MaSmoVaWin}. Thus one should consider also the possible existence of higher-order integrals. However, these are computationally very difficult to find. In this paper we have presented
two propositions, namely Propositions~\ref{lemma_minimally} and~\ref{lemma:HBmaxsuper} which can be used to construct three-dimensional maximally superintegrable systems with magnetic field out of two-dimensional scalar ones. In particular, Proposition~\ref{lemma:HBmaxsuper} states that a system with
\begin{gather*}
\vec B(\vec x)=(0, \gamma,0),\qquad \gamma\neq 0, \qquad W(\vec x)= V(x_2),
\end{gather*}
is maximally superintegrable if and only if the two-dimensional system with the Hamiltonian
\begin{gather*}
\mathcal K(\vec X,\vec P)=\frac12\big(P_1^2+P_2^2\big)+ \frac12 \gamma^2 X^2+ V(Y)
\end{gather*}
is superintegrable (where $V$ is the same function of a single variable).

Using Proposition~\ref{lemma:HBmaxsuper} we have arrived at an explicit example of maximally superintegrable system with
\begin{gather*}
\vec B(\vec x)=(0, \gamma,0), \qquad W(\vec x)= \frac{c}{x_2^2}+\frac{ m^2}{\ell^2} \gamma^2 x_2^2,\qquad \ell,m\in\mathbb{N}, \qquad c\in \mathbb{R},
\end{gather*}
cf.~\eqref{cagedOsc}, with three first-order integrals~\eqref{integralsB} and an additional integral coming from the integral of two-dimensional caged oscillator through the change of variables~\eqref{reducingtrB}.

Similarly, Proposition~\ref{lemma_minimally} led us to minimally superintegrable systems~\eqref{extcage2}
\begin{gather*}
 \vec B(\vec x) = 2 \left( \omega m_1 x_2-\frac{\beta_1}{x_2^3}, -\omega \ell_1 x_1+\frac{\alpha_1}{x_1^3}, 0\right), \\
 W(\vec x) = -\frac{\omega^2}{2} \big(\ell_1 x_1^2+m_1 x_2^2\big)^2+\omega \left( \ell_2 x_1^2+ m_2 x_2^2- \alpha_1 m_1 \frac{x_2^2}{x_1^2} - \beta_1 \ell_1 \frac{x_1^2}{x_2^2} \right) \\
 \hphantom{\vec B(\vec x) =}{} +\frac{\alpha_2}{x_1^2}+\frac{\beta_2}{x_2^2}-\frac{1}{2}\left(\frac{\alpha_1}{x_1^2}+\frac{\beta_1}{x_2^2}\right)^2, \qquad \frac{l_1}{m_1}=\frac{l_2}{m_2}=\frac{l^2}{m^2},\qquad l,m\in\mathbb{Z}.
\end{gather*}
Systems~I.b and~I.c of Section~\ref{Conclusion1} with $a_1\neq 0$ are special subcases of it when the integral~$X_3$ becomes second order one.

Another class of minimally superintegrable systems with
\begin{gather*}
\vec B(\vec x)=(a_1,a_2,0), \qquad W(\vec x)= v_{12} x_1^2 + v_{22} x_2^2 - \frac{1}{2} \left( a_2 x_1 + a_1 x_2\right)^2,\\
 \frac{v_{12}}{v_{22}} \in \mathbb{Q}, \qquad v_{12},v_{22}\neq 0
\end{gather*}
can be constructed out of anisotropic harmonic oscillator in two dimensions through the canonical transformation~\eqref{v12v22cantransf}.

Of course, more efficient and widely applicable tools for construction of higher-order superintegrable systems are needed. Given the recent rapid progress on a similar problem for scalar potentials~\cite{EscWinYur, Mar1,Mar2,MarSajWin2,PoWi} (see also references in~\cite{MiPoWin}) we hope that in foreseeable future we will be able to report on further development also in the case with magnetic field.

\appendix

\section[Solution of the determining equations in Case I for $\beta_{12}=\beta_{21}=0$]{Solution of the determining equations\\ in Case I for $\boldsymbol{\beta_{12}=\beta_{21}=0}$}\label{appendix:case1}

We consider here $\beta_{12}=\beta_{21}=0$. Thus $c_0$ and $c_1$ cannot both vanish, since by assumption the right-hand side of~\eqref{linearint} is not identically zero and, by~\eqref{coeffcond1}, $a_j=0$, $j=1,2$.
There are several subcases, according to whether equation~\eqref{eqtostart} is trivially satisfied or not.
For the reader's convenience, let us type it here again
\begin{gather}
 (\beta_{33} x_1+ \gamma_{23})u_1'(x_2)
+ (\beta_{33} x_2 - \gamma_{13}) u_2'(x_1)-c_1 =0.\label{eqtostartApp}
\end{gather}
We can have
\begin{itemize}\itemsep=0pt
\item[$(a)$] the above equation is not trivially satisfied for $u_2$ nor for $u_1$, thus $\beta_{33}\neq0$ or $\beta_{33}=0$ and both $\gamma_{13}$, $\gamma_{23}$ not vanishing;
\item[$(b)$] equation \eqref{eqtostartApp} is trivially satisfied for $u_2$ but not for $u_1$, thus $\beta_{33}=\gamma_{13}=0$ and $\gamma_{23}\neq0$;
\item[$(c)$] equation \eqref{eqtostartApp} is trivially satisfied for $u_1$ but not for $u_2$, thus $\beta_{33}=\gamma_{23}=0$ and $\gamma_{13}\neq0$;
\item[$(d)$] equation \eqref{eqtostartApp} is trivially satisfied for both~$u_j$, thus $c_1=\beta_{33}=\gamma_{23}=\gamma_{13}=0$.
\end{itemize}
Notice that Case $(b)$ can be reduced to Case $(c)$, and viceversa, by a canonical exchange of~$p_1$ with~$p_2$. Thus both cases are recovered by Appendix~\ref{CaseC1}. Case~$(a)$ is splitted for convenience into Appendices~\ref{CaseA1} and~\ref{CaseB1}, while Case~$(d)$ is treated in Appendix~\ref{CaseD1}.

\subsection[$\beta_{33}\neq0$]{$\boldsymbol{\beta_{33}\neq0}$}\label{CaseA1}

In this case, by translation in $x_1$ and $x_2$, we can set $\gamma_{13}=\gamma_{23}=0$.

By taking second derivatives of~\eqref{eqtostartApp} with respect to $x_1$ and $x_2$ and by equating them to zero, we find that
\begin{gather}\label{usolA}
u_1(x_2)= \frac{a_{12}}{2}x_2^2 + a_{11}x_2, \qquad u_2(x_1)=- \frac{a_{12}}{2}x_1^2 + a_{21}x_1,\qquad a_{ij}\in\mathbb{R}.
\end{gather}
By imposing then that~\eqref{usolA} solves~\eqref{eqtostartApp}, we obtain a polynomial expression in $x_1$ and $x_2$ which must vanish{\samepage
\begin{gather*}
c_1 - (a_{11}x_1 - a_{21}x_2) \beta_{33}=0.
\end{gather*}
We conclude $a_{11}=a_{21}=c_1=0$.}

Solving the remaining equations in \eqref{2ordcond} we find the functions $S_j$. Next we look at the first-order equations \eqref{1ordcond}, and in particular at the third one, for $m_{x_3}$. We proceed as above to solve it for $V_j$. By considering its second-order derivatives with respect to $x_1$ and $x_2$ and by equating them to zero, we find that also the functions $V_j$ must be second-order polynomials. Then we plug such solution back into the equation for $m_{x_3}$ and find a polynomial in $x_1$ and $x_2$ that must vanish. This implies $c_0=0$. Thus $\{X_3,p_3\}=0$. Nothing new can be found in this case.

\subsection[$\beta_{33}=0$, $\gamma_{13}\neq0$, $\gamma_{23}\neq0$]{$\boldsymbol{\beta_{33}=0}$, $\boldsymbol{\gamma_{13}\neq0}$, $\boldsymbol{\gamma_{23}\neq0}$}\label{CaseB1}

Proceeding as above we find from \eqref{eqtostartApp} that
\begin{gather*}
c_1= \left(a_{1} \gamma_{23}-a_{2} \gamma_{13}\right)
\end{gather*}
and
\begin{gather}\label{usolB}
u_1(x_2)= a_{1}x_2, \qquad u_2(x_1)= a_{2}x_1,\qquad a_{ij}\in\mathbb{R},
\end{gather}
where $|a_{1}|^2+|a_{2}|^2\neq 0$. The solution \eqref{usolB} implies that the magnetic field is constant.
By substituting \eqref{usolB} into the compatibility conditions \eqref{comps} for the magnetic field, they imply $\alpha_{33}=0$. The second-order equations~\eqref{2ordcond} can now be solved and give
\begin{gather*}%\label{ssolB}
S_1(x_1,x_2) = a_{2} \gamma_{13}x_1 +
 (S - a_{1} \gamma_{13} a_{2}\gamma_{23})x_2 + s_{11}, \qquad
S_2(x_1,x_2)= -S x_1 - a_{1} \gamma_{23} x_2 + s_{21}, \\
S_3(x_1,x_2) =
 \frac12\big(a_{1}\beta_{13}x_1^2 +a_{2}\beta_{23} x_2^2\big)- a_{2}\gamma_{1} x_2 + ( a_{2}\beta_{13} x_2+a_{1}\gamma_{1} ) x_1 ,\qquad S, s_{ij}\in\mathbb{R}\nonumber,
\end{gather*}
together with the condition
\begin{gather*}
\beta_{13} a_{2}=\beta_{23} a_{1}.
\end{gather*}
We now look at the first-order equations~\eqref{1ordcond}, starting with the equation for~$m_{x_3}$. Its first-order derivatives with respect to $x_1$ and $x_2$ imply that each $V_j$ is a second-order polynomial in its variable, $j=1,2$. This case is discussed in Section~\ref{sec:pol2}.

\subsection[$\beta_{33}=\gamma_{23}=0$, $\gamma_{13}\neq0$]{$\boldsymbol{\beta_{33}=\gamma_{23}=0}$, $\boldsymbol{\gamma_{13}\neq0}$}\label{CaseC1}

In this case equation \eqref{eqtostartApp} is trivially satisfied for $u_1$ and as an equation for $u_2$ implies
\begin{gather}\label{usolC}
u_2(x_1)= -\frac{c_1 x_1}{ \gamma_{13}}.
\end{gather}
By solving the second-order equations \eqref{2ordcond} for $s_j$, $j=1,2,3$,
we find
\begin{gather*}%\label{ssolC}
S_1(x_1,x_2) =-c_1 x_1 - S x_2 - \gamma_{13}u_1(x_2) + s_{11},\qquad S_2(x_1,x_2)=S x_1 + s_{21},\nonumber\\
S_3(x_1,x_2) =\frac{c_1 x_2 \big(2 \alpha_{33} x_1 x_2- 2 \beta_{13}x_1 - \frac12\beta_{23} x_2 + \gamma_{12}\big)}{ \gamma_{13}} \nonumber\\
\hphantom{S_3(x_1,x_2) =}{} + x_1 \left(x_1 \left(-\alpha_{33} x_2 + \frac12\beta_{13}\right) - \beta_{23} x_2 + \gamma_{12}\right) u_1'(x_2),
\end{gather*}
where $S, s_{ij}\in\mathbb{R}$ and $u_1$ has to solve the remaining compatibility conditions~\eqref{comps} (some are already satisfied by the conditions imposed on the constant $\alpha_{ij}$, $\beta_{ij}$, $\gamma_{ij}$ and~\eqref{usolC}):
\begin{align}
&\frac{6 c_1 \alpha_{33}}{\gamma_{13}}- \beta_{23} x_2 u_1^{(3)}(x_2)+ \gamma_{12}u_1^{(3)}(x_2)-4\beta_{23} u_1''(x_2)=0,\label{compsC}\\
& (2 \alpha_{33} x_2 -\beta_{13}) u_1''(x_2)+6 \alpha_{33}u_1'(x_2)=0\label{compsC-2}.
\end{align}
The second derivative with respect to $x_1$ of the third first-order equation in \eqref{1ordcond} implies
\begin{gather}\label{V1solC}
V_1(x_1)=v_{11} x_1+v_{12} x_1^2 ,\qquad v_{ij}\in\mathbb{R}.
\end{gather}
To proceed, we must solve the equations \eqref{compsC}--\eqref{compsC-2} for $u_1$. There can be several subcases.

\subsubsection[$\alpha_{33}=\beta_{13}=\beta_{23}=\gamma_{12}=0$]{$\boldsymbol{\alpha_{33}=\beta_{13}=\beta_{23}=\gamma_{12}=0}$}

In this case \eqref{compsC}--\eqref{compsC-2} are trivially satisfied. Let us consider the second set of compatibility conditions \eqref{compm} coming from the first-order equations, which simplify to
\begin{gather}
\frac{{c_1}^2}{\gamma_{13}}-2v_{12}\gamma_{13}-S u_1'(x_2)=0,\label{CmC-1}\\
c_1 \left( \frac{S}{\gamma_{13}} +u_1'(x_2) \right)-s_{21} u_1''(x_2)=0.\label{CmC-2}
\end{gather}
Let us first assume $u_1''\neq 0$. If $c_1=0$, then $S=v_{12}=s_{21}=0$ and $u_2(x_1)=0$. Equation~\eqref{0ordcond} implies that then $u_1$ is a constant, in contradiction with the assumption $u_1''\neq0$. Thus, let us consider the case $u_1''$ and $c_1$ both not zero. From equation \eqref{CmC-2} we obtain $S=0$ and
\begin{gather*}%\label{u2solC1}
u_1(x_2)=\frac{a_1 s_{21}}{c_1} {\rm e}^{\frac{c_1 x_2}{s_{21}}},\qquad a_1\in\mathbb{R}
\end{gather*}
together with
\begin{gather*}
v_{12}=\frac{c_1^2}{2 \gamma_{13}^2}
\end{gather*}
that we substitute into \eqref{V1solC}.

The solution of the compatibility constraints \eqref{CmC-1}--\eqref{CmC-2} assures that a solution of \eqref{1ordcond} for~$m$ exists. Thus, let us look at the zero-order equation \eqref{0ordcond}, which in this case reduces to
\begin{gather}\label{0C1}
c_1 v_{11} x_1 -\frac{a_1 s_{21} {\rm e}^{\frac{c_1 x_2}{s_{21}}} \big(4 v_{11} \gamma_{13}^2+c_1 s_{11}\big)}{ c_1\gamma_{13}}+s_{11} v_{11}+s_{21}V_2'(x_2)=0.
\end{gather}
Therefore $v_{11}=0$ and \eqref{V1solC} simplifies to
\begin{gather*}%\label{V1solC1}
V_1(x_1)=\frac{c_1^2 x_1^2}{2 \gamma_{13}^2}.
\end{gather*}
Let us notice that if $s_{21}=0$ we have constant magnetic field along the $x_2$ direction and arbit\-rary~$V_2$. Thus, we obtain the class of systems already studied in Section~\ref{sec:maximally_higher}. Otherwise, if $s_{21}$ is not zero, by solving~\eqref{0C1} we find
\begin{gather*}
V_2(x_2)=\frac{a_1 s_{11} s_{21} {\rm e}^{\frac{c_1 x_2}{s_{21}}}}{ c_1 \gamma_{13}}.
\end{gather*}
Thus, we arrive at the system determined by
\begin{gather} \label{BW31}
\vec B(\vec x)=\big( a {\mathrm{e}}^{b x_2}, c, 0\big),\qquad
 W(\vec x)= a \left( w +\frac{ c}{b} x_1 \right){\mathrm{e}}^{b x_2} -\frac{a^2}{2 b^2}{\mathrm{e}}^{2 b x_2},
\end{gather}
where we relabelled the integration constants as $a_1 = a$, $c_1 = -\gamma_{13} c$, $s_{11} = \gamma_{13} b w$, $s_{21} = -\frac{c \gamma_{13}}{b}$, with $a, b, c, w \in\mathbb{R}$ such that $a$, $b$, and $c$ are not vanishing.
The system~\eqref{BW31} is (at least) minimally superintegrable, with the integral $X_3$ as in \eqref{classintUEA}, where all the coefficients of the second-order terms are zero except~$\gamma_{13}$, and $S_j$ and $m_3$ of equation~\eqref{CaseIsm} are given by
\begin{gather*}%\label{ssolC1}
S_1(x_1, x_2) = \gamma_{13} \left(w b - {\mathrm{e}}^{b x_2} \frac{a}{b}+c x_1\right),\qquad S_2(x_1, x_2) = -\gamma_{13} \frac{c}{b},\qquad S_3(x_1, x_2)=0, \nonumber \\
m_3(\vec x) = \gamma_{13} c \left( {\mathrm{e}}^{b x_2} \frac{a}{b}-c x_1 -w b \right) x_3.
\end{gather*}

It remains to be considered the case in which $u_1''=0$, therefore $u_1(x_2)= a_1 x_2$, the magnetic field is constant. If $c_1=0$, from the zero-order equation~\eqref{0ordcond} we get that either~$V_2$ is a second-order polynomial (not of interest here since $V_1$ is already a second-order polynomial, too), or $S=s_{21}=v_{11}=0$ and $v_{12}=0$ from~\eqref{CmC-1}. This gives $V_1(x_1)=0$. From the first-order equations~\eqref{1ordcond} we find $c_0=0$. Nothing new here.

If $c_1$ is not zero, the compatibility condition~\eqref{CmC-1},~\eqref{CmC-2} together with the zero-order equation~\eqref{0ordcond} imply that either $V_1$ and $V_2$ are both second-order polynomials or~$V_1=0$ and the magnetic field is constant and aligned along the $x_2$-axis. Both these cases are of no interest in this section since they are considered elsewhere, in Section~\ref{sec:pol2} and Appendix~\ref{appendix:case2}.

\subsubsection[$\alpha_{33}=0$, $\beta_{13}\neq0$]{$\boldsymbol{\alpha_{33}=0}$, $\boldsymbol{\beta_{13}\neq0}$}
By translation in $x_1$ we can assume $\gamma_{12}=0$.

The compatibility condition \eqref{compsC} reduces to
\begin{gather}
-\beta_{23} \big( x_2 u_1^{(3)}(x_2)+4 u_1''(x_2)\big)=0,\nonumber \\ %\label{CsC1-1} \\
\beta_{13}u_1''(x_2)=0.\label{CsC1-2}
\end{gather}
Since $\beta_{13}$ is not zero, \eqref{CsC1-2} gives
\begin{gather*}%\label{u1solC1}
u_1(x_2)=a_1 x_2,\qquad a_1\in\mathbb{R}.
\end{gather*}
The compatibility conditions~\eqref{compm} for $m_3$ imply that $V_2$ is a second-order polynomial. This case is of no interest here and it is studied in Section~\ref{sec:pol2}.

\subsubsection[$\alpha_{33}=0$, $\beta_{13}=0$, $\beta_{23}\neq0$]{$\boldsymbol{\alpha_{33}=0}$, $\boldsymbol{\beta_{13}=0}$, $\boldsymbol{\beta_{23}\neq0}$}

Still, by translation in $x_1$, we can set $\gamma_{12}=0$.
Equations \eqref{2ordcond} imply
\begin{gather*}
u_1(x_2)=\frac{a_2}{6 x_2^2}.
\end{gather*}
In order to have nonvanishing magnetic field and nonvanishing Poisson bracket in~\eqref{linearint0}, we find from equations~\eqref{1ordcond} and~\eqref{0ordcond} that we must have $S=0$, $a_2=0$ and $c_0 = \frac{c_1 s_{11}}{\gamma_{13}}$. Looking for a~solution of~\eqref{0ordcond} for $V_2$ not second or lower-order polynomial we find
\begin{gather*}
\vec B(\vec x)=\left(0,-\frac{c_1}{\gamma_{13}},0\right), \qquad
V_1(x_1) = \frac{c_1^2 x_1^2}{2 \gamma_{13}^2}, \qquad V_2(x_2) = \frac{c_1^2 x_2^2}{8 \gamma_{13}^2}-\frac{v_{21}}{2 x_2^2},
\end{gather*}
which is the already known system~\eqref{1/x2potential}.

\subsubsection[$\alpha_{33}=\beta_{13}=\beta_{23}=0,$ $\gamma_{12}\neq0$]{$\boldsymbol{\alpha_{33}=\beta_{13}=\beta_{23}=0,$ $\gamma_{12}\neq0}$}

The compatibilities \eqref{compsC}--\eqref{compsC-2} reduce to the sole equation
\begin{gather*}
\gamma_{12}u_1^{(3)}(x_2)=0,
\end{gather*}
therefore
\begin{gather*}%\label{u1solC3}
u_1(x_2)=a_{1}x_2+ a_{2}x_2^2,\qquad a_j\in\mathbb{R}.
\end{gather*}
The compatibility conditions \eqref{compm}, together with the zero-order equation, imply that either the magnetic field is constant and $V_2$ is a second-order polynomial, or $V_1(x_1)=u_2(x_1)=0$ and $X_2$ reduces to a first-order integral. Both these cases are of no interest here.

\subsubsection[$\alpha_{33}\neq0$]{$\boldsymbol{\alpha_{33}\neq0}$}

By translation in $x_1$ and $x_2$ we can set $\beta_{13}=\beta_{23}=0$. Thus, \eqref{compsC-2} gives
\begin{gather*}
u_1(x_2)=\frac{a}{2 x_2^2},\qquad a\in \mathbb{R}.
\end{gather*}
By plugging this solution into \eqref{2ordcond} we obtain that it must be $c_1=0$ and therefore by \eqref{compsC} $\gamma_{12}=0$. By looking for a solution of the equations~\eqref{compm} and~\eqref{1ordcond} we find that also $c_0=0$, i.e., $\{ X,p_3\}=0$, and this case is of no interest here.

\subsection[$\beta_{33}=\gamma_{23}=\gamma_{13}=c_1=0$]{$\boldsymbol{\beta_{33}=\gamma_{23}=\gamma_{13}=c_1=0}$}\label{CaseD1}

In this case equation \eqref{eqtostartApp} is trivially satisfied for both $u_j$. The remaining second-order equations give for~$S_j$:
\begin{gather*}
S_1(x_1,x_2) =S x_2+ s_{11},\qquad S_2(x_1,x_2)=-S x_1+ s_{21},\nonumber\\
S_3(x_1,x_2) =-\frac{x_1}{2} (2 \alpha_{33}x_1 x_2 -\beta_{13}x_1 +2\beta_{23} x_2-2\gamma_{12})u_1'(x_2)\nonumber\\
\hphantom{S_3(x_1,x_2) =}{} -2 x_2(\alpha_{33}x_2-\beta_{13}) u_2(x_1)+S_{31}(x_2),\qquad S, s_{ij} \in\mathbb{R}, %\label{ssolD}
\end{gather*}
where $S_{31}(x_2)$ must solve
\begin{gather*}
\left(\alpha_{33} x_1^2 x_2-\frac{\beta_{13}}{2}x_1^2 +\beta_{23}x_1 x_2 - \gamma_{12}x_1\right) u_1''(x_2)
+3 x_1 (\alpha_{33}x_1 +\beta_{23})u_1'(x_2)\\
\qquad{} -
 ( (\beta_{13}-2\alpha_{33} x_2 ) x_1-\beta_{23} x_2+\gamma_{12})u_2'(x_1) \\
\qquad {} +2 (2\alpha_{33} x_2 -\beta_{13})u_2(x_1)-S_{31}'(x_2) = 0
\end{gather*}
and $u_j$ satisfy the compatibilities \eqref{comps}, that in this case reduce to
\begin{gather}
(2\alpha_{33} x_1 x_2-\beta_{13}x_1 +\beta_{23} x_2-\gamma_{12})u_1'''(x_2)
-4 (2 \alpha_{33}x_1 +\beta_{23})u_1''(x_2) \nonumber\\
\qquad{} +(2\alpha_{33}x_1 +\beta_{23})u_2''(x_1)+6 \alpha_{33} u_2'(x_1) = 0,\label{compD-1}\\
6\alpha_{33} u_1'(x_2)+(2\alpha_{33}x_2-\beta_{13})(u_1'(x_2)+ 8 u_2'(x_1))\nonumber \\
\qquad {}-(2\alpha_{33}x_1 x_2 -\beta_{13}x_1+\beta_{23}x_2-\gamma_{12})u_2'''(x_1) = 0.
\label{compD-2}
\end{gather}
As in the above Appendix~\ref{CaseC1}, to solve \eqref{compD-1}--\eqref{compD-2} we have to distinguish several subcases.

\subsubsection[$\alpha_{33}\neq0$]{$\boldsymbol{\alpha_{33}\neq0}$}
By translation in $x_1$ and $x_2$ we can eliminate $\beta_{13}$ and $\beta_{23}$. By taking the third-order derivatives~$\partial ^2_{x_1}\partial _{x_2}$, $\partial ^2_{x_2}\partial _{x_1}$ of~\eqref{compD-1} and~\eqref{compD-2}, respectively we get
\begin{gather*}
u_2(x_1) =a_{22} x_1^2-\frac{a_{23}}{x_1^2}+a_{21} x_1,\qquad
u_1(x_2) =a_{12} x_2^2-\frac{a_{13}}{x_2^2}+a_{11} x_2,\qquad a_{ij}\in\mathbb{R}.
\end{gather*}
By plugging the above solution into the third condition in \eqref{1ordcond} we see that, as a polynomial in~$x_1$ and~$x_2$, it can be vanishing only if~$c_0=0$. Since also $c_1=0$, there is nothing new here.

\subsubsection[$\alpha_{33}=0$, $\beta_{13}\neq0$]{$\boldsymbol{\alpha_{33}=0}$, $\boldsymbol{\beta_{13}\neq0}$}

By translation in $x_1$ we can assume $\gamma_{12}=0$. By considering the first-order derivative of~\eqref{compD-1} with respect to $x_2$ and of~\eqref{compD-2} with respect to~$x_1$, we find
\begin{gather}
u_1(x_2) =a_{11} x_2+ a_{12} x_2^2+ a_{13} x_2^3,\qquad
u_2(x_1) = a_{21} x_1+ a_{22} x_1^2+ \frac{a_{23}}{x_1^2},\qquad a_{ij}\in\mathbb{R}. \label{usolD-1}
\end{gather}
We plug \eqref{usolD-1} into \eqref{compD-1}--\eqref{compD-2} and we obtain the conditions $a_{23}=a_{13}=0$ and
\begin{gather*}
a_{22} \beta_{23}=0.
\end{gather*}
If $\beta_{23}\neq0$, then $a_{22}=0$. The equation for $m_{x_3}$ in~\eqref{1ordcond}, together with~\eqref{compm}, implies that $V_j$ are second-order polynomials and $a_{12}=0$, i.e., the magnetic field is constant. This case is of no interest here, cf.\ Section~\ref{sec:pol2}.

If $\beta_{23}=0$ we again consider the third equation in~\eqref{1ordcond}, together with~\eqref{compm} and their first order and second order mixed derivatives with respect to $x_1$ and $x_2$. If $a_{22}$ is not vanishing, we can find a solution for both $V_j$ which is not a second-order polynomial (or lower) but only assuming that $c_0=0$. Therefore nothing new arises in this case.

If $a_{22}=0$, we can find a solution for $V_j$ only if the magnetic field is vanishing, or $c_0=0$, or the solution is equivalent to the system~\eqref{1/x2potential}. Anyway, not of interest here (we recall that in Appendix~\ref{CaseD1} we have $c_1=0$).

\subsubsection[$\alpha_{33}=\beta_{13}=0$, $\beta_{23}\neq 0$]{$\boldsymbol{\alpha_{33}=\beta_{13}=0}$, $\boldsymbol{\beta_{23}\neq 0}$}

We can eliminate $\gamma_{12}$ by translation in $x_2$, thus setting $\gamma_{12}=0$.

The compatibility conditions \eqref{compD-1}, \eqref{compD-2} imply that
\begin{gather*}
u_1(x_2) = a_{12} x_2^2+\frac{a_{11}}{x_2^2}, \qquad u_2(x_1) = -4 a_{12} x_1^2+a_{21} x_1, \qquad a_{ij}\in\mathbb{R}.
\end{gather*}
By imposing that also the remaining equations~\eqref{2ordcond},~\eqref{1ordcond} and~\eqref{0ordcond} are satisfied we find that either $c_0=0$ which implies $\{ X,p_3 \}=0$ in contradiction with our assumption, or we find $ a_{11} =a_{12} =0$ and
\begin{gather*}
V_1(x_1) = \frac{1}{2} a_{21}^2 x_1^2, \qquad V_2(x_2) = \frac{a_{21}^2}{8} x_2^2 -\frac{v_{21}}{2 x_2^2},
\end{gather*} i.e., a system equivalent to~\eqref{1/x2potential}.

\subsubsection[$\alpha_{33}=\beta_{13}=\beta_{23}=0$, $\gamma_{12}\neq0$]{$\boldsymbol{\alpha_{33}=\beta_{13}=\beta_{23}=0}$, $\boldsymbol{\gamma_{12}\neq0}$}

The compatibility conditions \eqref{compD-1}, \eqref{compD-2} imply that
\begin{gather*}
u_1(x_2)= a_{12} x_2^2+a_{11} x_2,\qquad
u_2(x_1)= - a_{12} x_1^2+a_{21} x_1, \qquad a_{ij}\in\mathbb{R}. %\label{usolD-3}
\end{gather*}
By imposing that also the remaining equations are satisfied we find that necessarily $a_{12}=0$ and
\begin{gather*}
 V_1(x_1) = \frac{1}{2} \big( a_{11}^2+ a_{21}^2\big) x_1^2+v_{12} x_1,\qquad
V_2(x_2) = \frac{1}{2} \big( a_{11}^2+ a_{21}^2\big) x_2^2 +\frac{a_{21} v_{12}}{a_{11}} x_2.
\end{gather*}
Thus this case was already studied in Section \ref{sec:pol2}.

\section[Solution of the determining equations in Case I for $\beta_{12}$, $\beta_{21}$ not both vanishing]{Solution of the determining equations\\ in Case I for $\boldsymbol{\beta_{12}}$, $\boldsymbol{\beta_{21}}$ not both vanishing}\label{appendix:case1-part2}

Here we distinguish between the cases
\begin{itemize}\itemsep=0pt
\item [$(a)$] $\beta_{12}$ and $\beta_{21}$ both not vanishing;
\item [$(b)$] $\beta_{12}\neq0$ and $\beta_{21}=0$. Notice that, by a canonical permutation of the variables, this case is equivalent to
 $\beta_{21}\neq0$ and $\beta_{12}=0$.
\end{itemize}
Case $(a)$ is treated in the following Appendix~\ref{appendix:case1-part21}. Case $(b)$ follows in Appendix~\ref{appendix:case1-part22}.

\subsection[$\beta_{12}\neq0$, $\beta_{21}\neq0$]{$\boldsymbol{\beta_{12}\neq0}$, $\boldsymbol{\beta_{21}\neq0}$}\label{appendix:case1-part21}

Since $\beta_{12}$ and $\beta_{21}$ are both nonvanishing, by translation in $x_1$ and $x_2$ we can set $\gamma_{13}=\gamma_{23}=0$.

Let us start from the compatibility conditions \eqref{comps}, that read
\begin{gather}
(\beta_{33}x_1 +\beta_{21} x_2)u_1'''(x_2)+4 \beta_{21} u_1''(x_2)=0,\nonumber\\
\beta_{33}(u_1''(x_2)+u_2''(x_1))=0,\nonumber\\
( 2\alpha_{33} x_1 x_2-\beta_{13} x_1 + \beta_{23}x_2- \gamma_{12}) u_1'''(x_2)+4 (2\alpha_{33} x_1+\beta_{23})u_1''(x_2)\nonumber\\
\qquad{}+ (2 \alpha_{33} x_1 +\beta_{23}) u_2''(x_1)+6\alpha_{33} u_2'(x_1) = 0,\nonumber\\
 (\beta_{12}x_1+\beta_{33}x_2)u_2'''(x_1)+4\beta_{12} u_2''(x_1)=0,\nonumber\\
\beta_{21} u_2''(x_1)+\beta_{12} u_1''(x_2)=0,\nonumber\\
 (2\alpha_{33} x_1 x_2- \beta_{13}x_1+ \beta_{23}x_2- \gamma_{12})u_2'''(x_1)-4 (\beta_{13}-2\alpha_{33}x_2)u_2''(x_1) \nonumber\\
\qquad{} - (\beta_{13}-2 \alpha_{33}x_2)u_1''(x_2)+6\alpha_{33} u_1'(x_2)=0.\label{ComegaAppB}
\end{gather}
The above equations could be trivially satisfied for $u_j$ or not, depending on the constants $\alpha_{ij}$, $\beta_{ij}$, $\gamma_{ij}$. This determines a splitting in the computation.

\subsubsection[$\beta_{33}\neq0$]{$\boldsymbol{\beta_{33}\neq0}$}\label{appB11}
The second equation in \eqref{ComegaAppB} implies

\begin{gather*}
u_1(x_2)=\frac{a_{12}}{2} x_2^2+ a_{11} x_2,\qquad u_2(x_1)=-\frac{a_{12}}{2} x_1^2+ a_{21} x_1,\qquad a_{ij}\in\mathbb{R}.
\end{gather*}
By imposing that the above solution satisfies the remaining compatibilities \eqref{ComegaAppB} and that the magnetic field does not vanish, we find that $a_{12}=\alpha_{33}=0$, therefore the magnetic field is constant. The third second-order equation simplifies to
\begin{gather*}%\label{32ordAppB}
( \beta_{33} a_{11}+3\beta_{12}a_{21})x_1+ (3\beta_{21}a_{11}+ \beta_{33}a_{21})x_2=c_1,
\end{gather*}
which must hold for all values of $x_1$ and $x_2$. Without loss of generality, we can assume that one component of the magnetic fields is not vanishing, e.g.,~$a_{11}$. Thus the above equation implies
\begin{gather}\label{cond00B}
\beta_{33}=-3\frac{\beta_{12}a_{21}}{a_{11}},\qquad \beta_{21}=\frac{a_{21}^2}{a_{11}^2}\beta_{12}, \qquad c_1=0.
\end{gather}
With this assumption the first and second-order equations \eqref{2ordcond} can be solved for $S_1$ and $S_2$. We plug the so found solution into the third equation in \eqref{1ordcond} and take its third-order derivatives~$\partial^2_{x_1}\partial_{x_2}$ and $\partial^2_{x_2}\partial_{x_1}$. In this way we obtain the condition
\begin{gather*}
\beta_{12} a_{21} V_j'''(x_j)=0,\qquad j=1,2.
\end{gather*}
Since the magnetic field is already constant, we do not consider here solutions for $V_j$ in the form of at most quadratic polynomials. Thus necessarily $a_{21}=0$. However, from~\eqref{cond00B} we have $\beta_{21}=0$, which violates our assumption for this subcase.

\subsubsection[$\beta_{33}=0$]{$\boldsymbol{\beta_{33}=0}$}
Equations \eqref{ComegaAppB} can be solved for nonvanishing magnetic field only if $\alpha_{33}=0$. In this case we find the solution
\begin{gather*}
 u_1(x_2)=a_1 x_2,\qquad u_2(x_1)=a_2 x_1,\qquad a_j\in\mathbb{R},
\end{gather*}
corresponding to constant magnetic field. The third second-order equation then reduces to
\begin{gather*}
 a_1\beta_{21} x_2+ a_2\beta_{12} x_1=c_1,
\end{gather*}
which must hold for all $x_1$, $x_2$. Since $\beta_{12}$ and $\beta_{21}$ are assumed to be not vanishing in this section, we conclude that there is no solution for nonvanishing magnetic field.

\subsection[$\beta_{12}\neq0$, $\beta_{21}=0$]{$\boldsymbol{\beta_{12}\neq0}$, $\boldsymbol{\beta_{21}=0}$}\label{appendix:case1-part22}

Again, we start by the compatibilities \eqref{ComegaAppB}, now simplified by the condition $\beta_{21}=0$.
We can proceed as above, considering first the case $\beta_{33}\neq0$.
Then, by translation in $x_1$ and $x_2$ we can assume $\gamma_{23}=\gamma_{13}=0$. We proceed as in Appendix~\ref{appB11} with the simplification $\beta_{21}=0$. Equation~\eqref{cond00B} implies that the magnetic field has to vanish.

Therefore, we continue in the following by assuming $\beta_{33}=0$. Since $\beta_{12}\neq0$, by translation in $x_1$ we can still set $\gamma_{13}=0$. The computation then splits into two major subcases, according to whether $\gamma_{23}=0$ or not.

\subsubsection[$\gamma_{23}\neq0$]{$\boldsymbol{\gamma_{23}\neq0}$}
The third second-order equation simplifies to
\begin{gather}
2\beta_{12} u_2(x_1) + \gamma_{23} u_1'(x_2)+ \beta_{12}x_1 u_2'(x_1) =c_1.\label{eqtostartApp2}
\end{gather}
By solving for $u_j$ we find
\begin{gather*}
u_1(x_2)=a_1 x_2,\qquad u_2(x_1) = \frac{a_2}{2 x_1^2}+\frac{c_1-a_1 \gamma_{23}}{2 \beta_{12}},\qquad a_j \in\mathbb{R}.
\end{gather*}
The remaining second-order equations can be solved for $S_j$ only under the condition $ a_1 \beta_{23} =a_2 \beta_{23} = a_2 \gamma_{12} = a_1 \alpha_{33} = 0$. We find
\begin{gather*}
S_1(x_1, x_2) = -s_{21} x_2-\frac{a_2 \beta_{12}}{x_1}+s_{12}, \\
S_2(x_1, x_2) = \frac{ a_1 \beta_{12}}{2} x_1^2-a_1 \gamma_{23} x_2 + s_{21} x_1 + \frac{a_2 \gamma_{23}}{2 x_1^2}+s_{22}, \\
 S_3(x_1, x_2) = \frac{\beta_{13} a_1}{2} x_1^2+\gamma_{12} a_1 x_1- \alpha_{33} a_2 \frac{x_2^2}{x_1^2}+\beta_{13} a_2 \frac{x_2}{x_1^2}.
\end{gather*}
Equations~\eqref{1ordcond} imply the following constraint on our integration constants: $a_2 s_{21}=0$.
The compatibility conditions \eqref{compm} can be solved for $V_j$ and together with~\eqref{0ordcond} imply $a_2=0$. In order to have nonvanishing magnetic field we thus must have $\alpha_{33}=\beta_{23}=0$. Equation~\eqref{0ordcond} further implies $s_{21}=s_{12}=0$ together with
\begin{gather*}
V_1(x_1) = \frac{1}{8} a_1^2 x_1^2-\frac{v_1}{2 x_1^2}, \\
V_2(x_2) = \frac{1}{2} a_1^2 x_2^2+\big(a_1^2 \gamma_{23}^2-a_1 c_1 \gamma_{23}-2 c_0 \beta_{12}\big) \frac{x_2}{2 \gamma_{23}\beta_{12}}+\frac{c_0 x_2}{\gamma_{23}},
\end{gather*}
i.e., we arrived at a system equivalent to~\eqref{1/x2potential} (after translation in $x_2$).

\subsubsection[$\gamma_{23}=0$]{$\boldsymbol{\gamma_{23}=0}$}
Equation \eqref{eqtostartApp2} does not contain $u_1$ anymore, while for $u_2$ implies
\begin{gather*}
u_2(x_1)=\frac{a_2}{x_1^2}+\frac{c_1}{2\beta_{12}}, \qquad a_2\in\mathbb{R}.
\end{gather*}
From the last but one equation in \eqref{ComegaAppB} we see that
\begin{gather*}
u_1(x_2)=a_1 x_2,\qquad a_1\in\mathbb{R},
\end{gather*}
while the remaining equations \eqref{ComegaAppB} read
\begin{gather}
a_2\beta_{23}=0, \qquad \label{COmegaAppB22}
\alpha_{33} a_1 x_1^5- 4 a_2(\beta_{23}x_2-\gamma_{12})=0,
\end{gather}
i.e., $a_2\beta_{23}=\alpha_{33} a_1= a_2 \gamma_{12}=0$.

The second-order equations \eqref{2ordcond} can be solved for $S_1$ and $S_2$. We find
\begin{gather*}
S_1(x_1,x_2)=S x_2-\frac{2 a_2 \beta_{12}}{x_1}+ s_{11},\qquad
S_2(x_1,x_2)=-S x_1 +\frac{1}{2} a_1 \beta_{12} x_1^2+s_{21},\qquad s_{ij}\in\mathbb{R}.
\end{gather*}
The remaining second-order equations for $S_3$ imply that $\beta_{23}=0$, otherwise they would imply that the magnetic field must vanish.

To proceed, we need to solve \eqref{COmegaAppB22}. We have to distinguish between the cases $a_2=0$ and $a_2\neq0$.

Let us start by assuming $a_2=0$. The compatibility conditions \eqref{compm} together with the zero-order equation \eqref{0ordcond} can be solved only if $\alpha_{33}=\gamma_{12}=c_1=0$ and $S=s_{11}=s_{21}=0$. However, the resulting system is equivalent to the already known maximally superintegrable system \eqref{1/x2potential}, up to a permutation of the canonical variables.

If $a_2\neq0$, then from \eqref{COmegaAppB22} we have $\gamma_{12}=0$ and $\alpha_{33} a_1=0$. Under these two conditions, also the remaining second-order equations \eqref{2ordcond} can be solved for $S_3$. We find
\begin{gather*}
S_3(x_1,x_2)=\frac{1}{2} a_1 \beta_{13} x_1^2 +2 a_2 \left( \beta_{13}\frac{x_2}{x_1^2} - \alpha_{33}\frac{x_2^2}{x_1^2}\right).
\end{gather*}
The compatibility conditions for the first-order equations \eqref{compm} can be solved for $V_1$. They admit a solution for $V_2$ only if $\alpha_{33}=\beta_{13}=0$ (and in this way also the remaining \eqref{COmegaAppB22} is satisfied). After solving for $V_1$ and imposing the previous condition, we see that they are satisfied for any~$V_2$.
To find $V_2$ we look at the zero-order equation, in which we impose all the conditions we obtained till now and the solutions found for $V_1$ and $S_j$. We obtain in this way a polynomial in $x_1$ (whose some coefficients contains equations for $V_2$) that must vanish. By collecting the different powers of $x_1$ and impose that they are all equal to zero, we arrive at the condition $a_2\beta_{12}=0$, which cannot be satisfied in the case we are considering here. Thus, no new system can be found.

\section{Solution of the determining equations for Case II}\label{appendix:case2}
Let us start by some preliminary considerations. By taking second-order derivatives with respect to $x_2$ of \eqref{1ordcond} and \eqref{0ordcond} we obtain the conditions
\begin{gather*}
s_1''(x_2)U_2''(x_1)=0 \qquad \text{and}\qquad
s_1''(x_2)W'(x_1)=0.
\end{gather*}
If $s_1''\neq0$, we therefore have $U_2''=W'=0$. As a consequence, new solutions can arise here only for $U_3'''\neq0$ (i.e., for nonconstant magnetic field). If so, \eqref{COmegaU2} imply $\beta_{33}=\beta_{31}=0$. However then the compatibility conditions \eqref{compm} cannot be solved for $s_1''\neq0$.

Thus, necessarily $s_1''=0$, which gives
\begin{gather} \label{Cs1}
s_1(x_2)=s_{11}+s_{12}x_2.
\end{gather}
Let us proceed by looking at \eqref{COmegaU2} in case $\beta_{33}\neq0$. Then $U_2'''=0$ and the third equation in~\eqref{COmegaU2} can be solved for~$U_3$. Then the remaining equations imply that the magnetic field and the potential $W$ are constant, which is not of interest here. Thus, necessarily $\beta_{33}=0$ and~\eqref{Cs1} holds.

\subsection[$\alpha_{33}\neq0$]{$\boldsymbol{\alpha_{33}\neq0}$}\label{app:C-system0}

By translation in both $x_1$ and $x_2$ we can set $\beta_{13}=\beta_{23}=0$.
The conditions \eqref{COmegaU2} and \eqref{2ordcond} can be solved. We find
\begin{gather*}
U_2(x_1)=\frac{a_1}{2x_1},\qquad\! U_3(x_1)=\frac{a_2}{2x_1},\qquad\! s_3(x_1)=0,\qquad\! s_2(x_1)=-s_{12}x_1\nonumber,\qquad\! a_1,a_2,s_{12}\in\mathbb{R},
\end{gather*}
together with $c_2=\beta_{21}=\beta_{31}=\gamma_{13}=\gamma_{12}=0$. This give the solution for~$S_j$, $j=1,2,3$. The compatibility conditions~\eqref{compm} can then be solved for $W$ and by imposing also \eqref{0ordcond} give
\begin{gather}\label{WSC0}
W(x_1)=-\frac{a}{8x_1^4}+\frac{w}{ x_1^2},\qquad a,w\in\mathbb{R},
\end{gather}
together with $s_{11}=s_{12}=\beta_{12}=a_2=0$, where $a\equiv a_1$. The magnetic field is
\begin{gather}\label{BSC0}
\vec B(\vec x)=\big(0, a x_1^{-3},0\big),\qquad a\in\mathbb{R}.
\end{gather}
The first-order equations \eqref{compm} give
\begin{gather*}
M(x_1,x_2)=\alpha_{33} \left( \frac{2w x_2^2}{x_1^2} -\frac{a^2 x_2^2}{2 x_1^4} \right)
\end{gather*}
with the condition $c_0=c_1=0$. This system is a special case of the systems in Table~\ref{table:2Dquad}, namely it can be expressed in the forms both $\mathcal{E}_{1}$ and $\mathcal{E}_{2}$. The integral constructed here is included in the ones coming from $\mathcal{E}_{1}$ and $\mathcal{E}_{2}$, i.e., the system is only minimally quadratically superintegrable.

From now on we continue our search by assuming $\alpha_{33}=0$.

\subsection[$\alpha_{33}=0$, $\beta_{23}\neq0$]{$\boldsymbol{\alpha_{33}=0}$, $\boldsymbol{\beta_{23}\neq0}$}
By translation in $x_2$ we can set $\gamma_{12}=0$. Then the second and fourth equation in \eqref{COmegaU2} imply
\begin{gather*}
U_2(x_1)=a_{11} x_1^2,\qquad U_3(x_1)=a_{21} x_1^2+ a_{22} x_1^3, \qquad a_{ij}\in\mathbb{R}.
\end{gather*}
From the compatibility conditions \eqref{compm} and the zero-order equation \eqref{0ordcond} we easily see that the only possibility is polynomial potential at most quadratic and $a_{22}=0$ (and $a_{21}a_{11}=0$), i.e., constant magnetic field. Thus, this is not of interest here.

\subsection[$\alpha_{33}=\beta_{23}=0$, $\beta_{31}\neq0$]{$\boldsymbol{\alpha_{33}=\beta_{23}=0}$, $\boldsymbol{\beta_{31}\neq0}$}\label{app:CaseII-b31not0}

Equation~\eqref{eqtostart2} reads
\begin{gather*}%\label{eqtostart2s}
(\beta_{12}x_1- \gamma_{13})U_2''(x_1)+2\beta_{12} U_2'(x_1)-\beta_{31} U_3'(x_1)-c_2=0,
\end{gather*}
which, integrated by $x_1$ and neglecting integration constants (which do not affect the magnetic field), gives
\begin{gather}\label{U3U2rel}
U_3(x_1)=\frac{1}{\beta_{31}}\big((\beta_{12}x_1- \gamma_{13})U_2'(x_1)+\beta_{12} U_2(x_1)+ c_2 x_1\big).
\end{gather}
Conditions \eqref{compm} together with \eqref{1ordcond} and \eqref{0ordcond} seen as a polynomial in $x_2$ imply that unless $s_{12}=0$ the magnetic field vanishes. Since this is of no interest here, we continue in the following by assuming $s_{12}=0$. From the second-order equations~\eqref{2ordcond}, considered as polynomials in~$x_2$, we see that necessarily $s_3(x_1)=s_2(x_1)=0$. We shall distinguish several subcases.

\subsubsection[$\beta_{12}=\gamma_{13}=0$]{$\boldsymbol{\beta_{12}=\gamma_{13}=0}$}

Let us start by the easiest case in which $\beta_{12}=\gamma_{13}=0$.

By \eqref{U3U2rel} we have here $U_3''=0$. Thus $U_2''$ cannot vanish, otherwise we have no magnetic field. From equations \eqref{2ordcond} we see that necessarily $\beta_{21}=0$. At this point, all \eqref{2ordcond} are solved, except for the condition
\begin{gather}\label{eqU200-last}
 (\beta_{13}x_1+\gamma_{12} )U_2'''(x_1)+ (3 \beta_{13}+\beta_{31} ) U_2''(x_1)=0.
\end{gather}
The zero-order equation reads
\begin{gather*}
\left(s_{11}+\frac{c_2\gamma_{12}}{\beta_{31}}\right)W'(x_1)=0.
\end{gather*}

For $s_{11}\neq-\frac{c_2\gamma_{12}}{\beta_{31}}$ the above condition implies vanishing potential $W(x_1)$. From~\eqref{1ordcond} we see that then also $U_2''$ is constant. Thus, this solution is of no interest here.

Thus, let us set $s_{11}=-\frac{c_2\gamma_{12}}{\beta_{31}}$. In this case the third equation in \eqref{1ordcond} is solved for $c_0=-\frac{c_1 c_2}{\beta_{31}}$. The remaining first-order equations can be solved for $M$ and give
\begin{gather*}%\label{Msol00}
M(x_1,x_2)= (\beta_{13}x_1+\gamma_{12}) x_2 W'(x_1)
\end{gather*}
for $W(x_1)$ satisfying
\begin{gather}\label{eqW00-last}
 \big( W''(x_1)+ U_2''(x_1)^2\big)( \beta_{13}x_1+\gamma_{12})+ (c_1-\beta_{31} U_2'(x_1) )U_2''(x_1)+3 \beta_{13} W'(x_1)=0.
\end{gather}
Recall that $U_2$ needs to satisfy~\eqref{eqU200-last}, whose solution depends on the constants involved there. We can have

{\bf (a) $\boldsymbol{\beta_{13}=0}$.}
If $\gamma_{12}\neq0$,~\eqref{eqU200-last} is solved by
\begin{gather*}
U_2(x_1)= \frac{a}{b^2} {\rm e}^{b x_1},\qquad a,b\in\mathbb{R},\qquad b=-\frac{\beta_{31}}{\gamma_{12}} \in \mathbb{R},
\end{gather*}
giving a new superintegrable system
\begin{gather}\label{BWSCa}
\vec B(\vec x)=\big(0,a {\rm e}^{b x_1},0\big),\qquad W(\vec x)= w x_1+ c e ^{b x_1}-\frac12 \frac{a^2}{b^2} {\rm e}^{2 bx_1},\qquad a,w\in \mathbb{R},
\end{gather}
where $c=-\frac{a c_1 \gamma_{12}}{\beta^2_{31}}\in \mathbb{R}$.

For $\gamma_{12}=0$, equation \eqref{eqU200-last} reduces to $\beta_{31} U_2''(x_1)=0$, implying $U_2''(x_1)=0$ and therefore no magnetic field.

{\bf (b) $\boldsymbol{\beta_{31}\neq-\beta_{13}\neq\frac12\beta_{31}}$, $\boldsymbol{\beta_{13}\neq0}$.}
By translation in $x_1$ we can set $\gamma_{12}=0$. After the substitution $\beta_{31}=- b \beta_{13}$, the solution of~\eqref{eqU200-last} in this case reads
\begin{gather}\label{USCb}
U_2(x_1)=\frac{a x_1^{b}}{b-1},\qquad a\in\mathbb{R},\qquad b\neq 0,1,2.
\end{gather}
By solving \eqref{eqW00-last} for $W$, we arrive at a new system determined by
\begin{gather}\label{BWSCb}
\vec B(\vec x)=\big(0,a (b-2) x_1^{b-3},0\big),\qquad W(\vec x)=-\frac{a^2 x_1^{2(b-2)}}{2}+ a (b-2) c x_1^{b-2}+ \frac{w}{x_1^2},
\end{gather}
where $a, b, w, c= \frac{c_1}{b (2-b) \beta_{13}} \in\mathbb{R}$, $b\neq0,1,2$.

Notice that in the limit $b\rightarrow 0$ we reduce~\eqref{BWSCb} to the system determined by~\eqref{WSC0} and~\eqref{BSC0}, that therefore can be seen as a special case of~\eqref{BWSCb} (which is already known to have two second-order integrals, with highest-order terms $\big(l_3^A\big)^2$ and $ p_1^Al_3^A$, that are however dependent once also the Cartesian integrals~\eqref{CartInt2} and the Hamiltonian are taken into account).

{\bf (c) $\boldsymbol{\beta_{13}=-\frac12\beta_{31}}$.}
Since in this case $\beta_{13}=-\frac12\beta_{31}\neq0$, by translation in~$x_1$ we can still set $\gamma_{12}=0$. Equation~\eqref{eqU200-last} has solution
\begin{gather*}
U_2(x_1)=a x_1 \ln |x_1|,\qquad a\in\mathbb{R},
\end{gather*}
that gives the new system
\begin{gather}\label{BWSCc}
\vec B(\vec x)=\big(0,a x_1^{-1},0\big) ,\qquad W(\vec x)=-\frac12 a^2 (\ln |x_1|)^2 + b \ln |x_1|+ \frac{w}{x_1^2},\qquad a,w\in \mathbb{R},
\end{gather}
where $b=\frac{a c_1}{\beta_{31}}-a^2$.

{\bf (d) $\boldsymbol{\beta_{13}=-\beta_{31}}$.}
As above, we can still set $\gamma_{12}=0$ by translation. Equation~\eqref{eqU200-last} admits the solution
\begin{gather}\label{USCd}
U_2(x_1)=-a\ln |x_1|,\qquad a\in\mathbb{R}.
\end{gather}
By denoting $c=-\frac{c_1 a}{ \beta_{13}}$, from \eqref{eqW00-last} and \eqref{USCd} we obtain
\begin{gather}\label{BWSCd}
\vec B(\vec x)=\big(0,a x_1^{-2},0\big),\qquad W(\vec x)=\frac{c}{x_1}+ \frac{w}{x_1^2},\qquad a,c,w\in \mathbb{R}.
\end{gather}
Thus, though \eqref{USCb} is singular for $b=1$ (and we obtain a different solution~\eqref{USCd} for~\eqref{eqU200-last} if $b=1$), indeed such a singularity does not appear in the solution for $\vec B$ and~$W$ (it is lost by differentiating) and the system~\eqref{BWSCb} becomes~\eqref{BWSCd} for $b=1$.

\subsubsection[$\gamma_{13}\neq0$, $\beta_{12}=0$]{$\boldsymbol{\gamma_{13}\neq0}$, $\boldsymbol{\beta_{12}=0}$}

In this case it is convenient to start by looking at the zeroth-order equation~\eqref{0ordcond}, that simplifies to
\begin{gather} \label{C33eq}
 \big(\gamma_{13}\big((\beta_{13}x_1+\gamma_{12}) U_2''(x_1)- (\beta_{13}-\beta_{31})\big)U_2'(x_1) + s_{11}\beta_{31}+c_2\gamma_{12}\big)W'(x_1)=0.
\end{gather}
For $W'\neq0$, we solve the above equation for $U_2$. Once we plug the so found solution into the remaining equations we see that there can be several subcases according to different values of the constants involved. However, in most cases, the solution for~$U_2$ is polynomial and at most of first order. By~\eqref{U3U2rel} also $U_3$ results into a polynomial of order at most one. Thus in these cases we have vanishing magnetic field. The only exception is the solution
\begin{gather}
U_2 \left(x_1\right) = \frac{2 \gamma_{12} a ^2 \big(b_2^2+\frac{1}{2}\big) { e}^{b_1 x_1}-\big((2 b_1 b_2 s_{11}+c_1) a +w_2 b_1^2 \gamma_{12}\big) b_1 x_1}{b_1^2 \gamma_{12} a \big(2 b_2^2+1\big)}, \nonumber \\
\vec B(\vec x) = ( 0, 1, b_2 ) \cdot a {\textrm e}^{b_1 x_1},\nonumber \\ \nonumber
W(\vec x) = -\frac{\big( b_2^2 +1\big)}{2 b_1^2} a ^2 {\textrm e}^{2 b_1 x_1}+w_2 {\textrm e}^{b_1 x_1}+x_1 w_1,\nonumber \\
M (x_1, x_2) = -\frac{\gamma_{12}}{b_1} \big(a ^2 \big(b_2^2+1\big) {\textrm e}^{2 b_1 x_1}-w_2 b_1^2 {\textrm e}^{b_1 x_1}-w_1 b_1\big) x_2,\label{BWC32}
\end{gather}
where
$s_{12} = 0$, $\beta_{13} = 0$, $\beta_{21} = b_1 b_2 \gamma_{12}$, $\beta_{31} = -b_1 \gamma_{12}$, $\gamma_{13} = b_2 \gamma_{12}$, $c_0 = \gamma_{12} b_2 w_1+\frac{b_1^3 \gamma_{12} w_2^2 b_2^3}{a ^2 (2 b_2^2+1)^2}-\frac{b_1 b_2 (2 b_1 b_2 s_{11}+c_1) w_2}{a (2 b_2^2+1)^2}+\frac{(b_1 s_{11}-b_2 c_1) (b_1 b_2 s_{11}+b_2^2 c_1+c_1)}{b_1 \gamma_{12} (2 b_2^2+1)^2}$ and $c_2 = \frac{a b_1 s_{11}-w_2 b_1^2 b_2 \gamma_{12}-a b_2 c_1}{a \left(2 b_2^2+1\right)}$.
By rotation of the coordinate frame around the $x_1$-axis, which preserves the existence of the Cartesian integrals $p_2$ and $p_3$, the system~\eqref{BWC32} can be brought to the form~\eqref{BWSCa}.

For $W'=0$ the equation~\eqref{C33eq} does not give any condition on $U_2$, that is therefore constrained only by the remaining second and first-order equations~\eqref{2ordcond}, \eqref{1ordcond}. However, also in this case we always arrive to a solution for~$U_2$ at most linear in~$x_1$, except for
$U_2=\frac{1}{b^2}(a_1\sin(b x_1)+ a_2 \cos(b x_1))$
that leads to the superintegrable system already found in~\cite{MSW}, whose magnetic field has components
\begin{gather*}
B_1=0,\qquad B_2=a_1\sin(b x_1)+ a_2 \cos(b x_1),\qquad B_3=a_2\sin(b x_1)- a_1 \cos(b x_1),
\end{gather*}
where $a_1,a_2\in \mathbb{R}$ and $b=\frac{\beta_{31}}{\gamma_{13}}$ (which must for this case be also equal to $b=\frac{\beta_{21}}{\gamma_{12}}$).

\subsubsection[$\beta_{12}\neq0$]{$\boldsymbol{\beta_{12}\neq0}$}
By translation in $x_1$, let us set $\gamma_{13}=0$. We can proceed as above and, after a long and tedious computation, arrive at two solutions
\begin{itemize}\itemsep=0pt
\item if $W$ is not identically zero, we have
\begin{gather}
U_2(x_1) = \frac{1}{ b_1-1} \left(a x_1^{b_1-1}+\frac{b_1 s_{11}}{ b_2 \beta_{13}}\right), \nonumber \\
\vec B(\vec x) = ( 0, 1, b_2 ) \cdot a (b_1-2) x_1^{b_1-3},\nonumber\\
W(\vec x) = -\frac{a^2 }{2} \big(b_2^2+1\big) x_1^{2(b_1-2)}+w_2 x_1^{b_1-2}+\frac{w_1}{x_1^2},\nonumber\\
M(x_1, x_2) = \beta_{13} x_2 \left( (b_1-2) \big( w_2 x_1^{b_1} - a^2 \big(b_2^2+1\big) x_1^{2 (b_1-2)} \big) -2 \frac{w_1}{x_1^2} \right),\label{BWC33}
\end{gather}
where $c_0 = -\frac{c_2}{a} w_2-\frac{(b_1-1) b_2 c_2^2}{b_1^2 \beta_{13}}$, $c_1 = -2 (b_1-1) \frac{b_2}{b_1} c_2 -\frac{b_1}{a} \beta_{13} w_2$, $s_{12} = 0$, $\beta_{21} = b_1 b_2 \beta_{13}$, $\beta_{12} = -b_2 \beta_{13}$, $\beta_{31} = -b_1 \beta_{13}$ and $\gamma_{12} = 0$. By rotation of the coordinate frame around the $x_1$-axis, which preserves the existence of the Cartesian integrals $p_2$ and $p_3$, the system~\eqref{BWC33} can be brought to the form~\eqref{BWSCb}.
\item For $W(\vec x)=0$ we find
\begin{gather*}
U_2(x_1) = - a \ln|x_1|, \qquad
\vec B(\vec x) = ( 0, 1,b) \cdot \frac{a}{x_1^2},\qquad W(\vec x)=0, \qquad M(x_1,x_2)=0,
\end{gather*}
where $c_0 = 0$, $c_1 = 0$, $s_{11} = - a b \beta_{31}$, $\beta_{13} = -\beta_{31}$, $\beta_{21} = -b \beta_{31}$ and $\gamma_{12} = 0$. However, this solution can be obtained as a limiting case of~\eqref{BWC33} with $b_1=1$, $b_2=b$, $w_2=0$ and $w_1=\frac{a^2 (b_2^2+1)}{2}$.
\end{itemize}

\subsection[$\alpha_{33}=\beta_{23}=\beta_{31}=0$]{$\boldsymbol{\alpha_{33}=\beta_{23}=\beta_{31}=0}$}\label{app:CaseII-b310}

Equation \eqref{eqtostart2} now reads
\begin{gather}\label{eqtostart2ss}
(\beta_{12}x_1- \gamma_{13})U_2''(x_1)+2\beta_{12} U_2'(x_1)-c_2=0.
\end{gather}
Thus, it does not imply any relationship between $U_2$ and $U_3$. It only gives a condition on~$U_2$ that could be trivially satisfied or not, according to the values of the constants involved.
By the same argument used in Appendix~\ref{app:CaseII-b31not0},
we see that also in this case $s_{12}=0$ and $s_3(x_1)=s_2(x_1)=0$.

\subsubsection[$\beta_{21}\neq0$]{$\boldsymbol{\beta_{21}\neq0}$}

Let us look at the situation in which $\beta_{21}\neq0$ and recall that we are looking for an integral~$X$ of the form~\eqref{Comm2}. Its Poisson bracket with $X_1=p_2$, i.e., $\{X, p_2 \}$,
is again an integral, by assumption expressible in terms of the known integrals, similarly to equation~\eqref{Comm2}. However, these two expressions can be interchanged under the permutation of~$x_2$ and $x_3$; thus we can reduce the considered problem into an already discussed one. We have that
\begin{gather*}
(p_1,p_2,p_3,l_1,l_2,l_3)\rightarrow(p_1,p_3,p_2,-l_1,-l_3,-l_2),
\end{gather*}
while the system \eqref{Omega_sep2} transforms into
\begin{gather}\label{p2p3permutedsystem}
W(x_1)\rightarrow W(x_1),\qquad (u_2,u_3)\rightarrow(-u_3,u_2).
\end{gather}
Therefore, all the conditions imposed in this section (including $\beta_{21}\neq0$) on the coefficients of the second-order terms in the integral change into $\alpha_{22}=\beta_{32}=\beta_{21}=\beta_{22}=0$, $\beta_{31}\neq0$ and \eqref{alphaComm2} turns into $a_0=-\beta_{13}$, $a_1=0$, $a_2=\beta_{31}$, $a_3=\beta_{21}$ and
\begin{gather*}
 \alpha_{11}=\alpha_{12}=\alpha_{13}=\alpha_{23}=\alpha_{33}=\beta_{11}=\beta_{33}=\beta_{23}=0.
\end{gather*}
Thus we recover all the conditions imposed on our parameters in Appendix~\ref{app:CaseII-b31not0}. By~\eqref{p2p3permutedsystem} any system belonging into this section would transform into some system already found in Appendix~\ref{app:CaseII-b31not0}.
Therefore in the following we can assume that $\beta_{21}=0$ {without lost of generality}.

We distinguish several subcases depending on whether~\eqref{eqtostart2ss} is trivially satisfied or not.

\subsubsection[$\beta_{21}=0$, $\beta_{12}\neq0$]{$\boldsymbol{\beta_{21}=0}$, $\boldsymbol{\beta_{12}\neq0}$}

By translation in $x_1$ we can set $\gamma_{13}=0$. Equation \eqref{eqtostart2ss} gives
\begin{gather*}
U_2(x_1)=\frac{a_1}{2 x_1}+ \frac{c_2}{2\beta_{12}} x_1,\qquad a_1\in\mathbb{R}.
\end{gather*}
We look at the remaining second-order equations \eqref{2ordcond},
that read
\begin{gather*}
(\beta_{13}x_1+ \gamma_{12})U_3'''(x_1) + 3 \beta_{13} U_3''(x_1) = 0,\qquad
 \beta_{12} x_1^5 U_3'''(x_1)+ 3 \beta_{12} x_1^4 U_3''(x_1) - 3 \gamma_{12} a_1 = 0.
\end{gather*}

If $\beta_{13}=0$ and $\gamma_{12}\neq0$ they have a solution only if $a_1=0$ and $U_3=0$, i.e., the magnetic field vanishes.

If $\beta_{13}=\gamma_{12}=0$, the above equations admit solution for
\begin{gather*}
U_3(x_1)=\frac{a_2}{x_1}.
\end{gather*}
The zeroth-order equation \eqref{0ordcond} has solution for $W(x_1)=0$.
If $\beta_{12}\neq0$ the remaining equations~\eqref{1ordcond} lead to the solution
\begin{gather}
U_2(x_1) = 0, \qquad U_3(x_1) = \frac{a}{2 x_1}, \qquad \vec B(\vec x) = \left( 0, 0, \frac{a}{x_1^3} \right), \nonumber\\
W(\vec x) = -\frac{a b \ln|x_1|}{x_1^2}-\frac{a^2}{8 x_1^4}+\frac{w}{x_1^2},\label{C42UBW}
\end{gather}
where $a_1 = 0$, $c_0 = 0$, $c_1 = 2 \beta_{12} b$, $c_2=0$, $s_{11} = 0$, $\beta_{13} = 0$, $\gamma_{12} = 0$ and $M(x_1, x_2) = 0$. This system is new when $b\neq 0$, otherwise it can be turned into a subcase of~\eqref{BWSCb} by permutation of~$x_2$ and~$x_3$.

If $\beta_{13}\neq0$ we can set $\gamma_{12}=0$. We have a solution for $U_3$ given by
\begin{gather*}
U_3(x_1)=\frac{4 a_2}{\beta_{13}x_1},\qquad b\in\mathbb{R}.
\end{gather*}
Equations \eqref{1ordcond} and \eqref{0ordcond} can be solved for $W$ and $M$ under the conditions $s_{11}=0$ and $a_2= -\frac{a_1 \beta_{12}}{8}
 $.
We find
\begin{gather*}
M(x_1, x_2) = \beta_{13} \big(b^2+1\big) a^2\frac{x_2}{2 x_1^4}- \beta_{13} c \frac{2 \ln|x_1|-1}{x_1^2}-2 \beta_{13} w\frac{x_2}{x_1^2}
\end{gather*}
and
\begin{gather}\label{WSCd}
W(x_1)=-\frac{\big(b^2+1\big) a^2}{8 x_1^4}+\frac{c\ln |x_1|+ w}{x_1^2},\qquad w\in\mathbb{R},
\end{gather}
where $b=-\frac{\beta_{12}}{\beta_{13}}\neq0$, $a=a_1$, $ c=\frac{ a c_1}{2 \beta_{13}} $ and by using the physically irrelevant shift of the potential by an additive constant we set $c_0=c_2=0$.

The magnetic field reads
\begin{gather}\label{BSCd}
\vec B(\vec x)=(0,1, b )\cdot a x_1^{-3},\qquad a ,b\in\mathbb{R},\qquad b\neq0.
\end{gather}
However, the reference frame can be rotated around $x_1$ without affecting the fact that~$p_2$ and~$p_3$ are integrals. Such a rotation brings the magnetic field~\eqref{BSCd} and the potential~\eqref{WSCd} and the integral into the form~\eqref{C42UBW}.

\subsubsection[$\beta_{21}=\beta_{12}=0$, $\gamma_{13}\neq0$]{$\boldsymbol{\beta_{21}=\beta_{12}=0}$, $\boldsymbol{\gamma_{13}\neq0}$}

From equation \eqref{eqtostart2ss} we have
\begin{gather*}%\label{U2SolC41}
U_2(x_1)=-\frac{ c_2 x_1^2}{2\gamma_{13}}.
\end{gather*}
From the second and zeroth-order equations we get
\begin{gather*}
U_3(x_1)= a x_1^2,\qquad a\in\mathbb{R}.
\end{gather*}
The first-order equations can be solved for $W$ and give a polynomial solution at most quadratic. Since the magnetic field is constant, nothing of interest can be found here.

\subsubsection[$\beta_{21}=\beta_{12}=\gamma_{13}=0$]{$\boldsymbol{\beta_{21}=\beta_{12}=\gamma_{13}=0}$}

Equation \eqref{eqtostart2ss} implies $c_2=0$ and it is trivially satisfied for $U_2$. The remaining second-order equations read
\begin{gather*}
\beta_{13}\big( 3 U_j''(x_1)+ x_1 U_j'''(x_1) \big)+ \gamma_{12} U_j'''(x_1)=0,\qquad j=2,3.
\end{gather*}
If $\beta_{13}=0$ we easily conclude that the only possibility is constant magnetic field and vanishing potential.
For $\beta_{13}\neq0$ we can set $\gamma_{12}=0$ by translation in $x_1$. The above equations imply
\begin{gather*}
U_j(x_1)=\frac{a_j}{2 x_1},\qquad j=2,3.
\end{gather*}
From the first-order equations, seen as polynomials in $x_2$, we get the conditions
\begin{gather*}
\frac{ a_2^2}{x_1^5}+\frac{a_3^2}{x_1^5}-2 W'(x_1)=0, \qquad
 a_2( 2 a_3\beta_{13}+ 3 s_{11} x_1)+c_1 a_3 x_1^2=0,
\end{gather*}
together with two differential equations for $M$.
From the second equation above we see that we have two possibilities for nonvanishing magnetic field: $a_2=c_1=0$ or $a_3=s_{11}=0$.
In the first case, when imposing also the remaining zero-order equation \eqref{0ordcond} we get that necessarily $a_3=0$, i.e., the magnetic field vanishes. In the second case we have the solution
\begin{gather*}
W(x_1)=\frac{c\ln |x_1| + w}{x_1^2}-\frac{a^2}{8 x_1^4},\qquad a,c, w\in\mathbb{R},
\end{gather*}
with $c= \frac{a c_1}{2\beta_{13}}$, $a=a_2$. The magnetic field reads
\begin{gather*}
\vec B(\vec x)=\big(0, a x_1^{-3},0\big),\qquad a\in\mathbb{R}.
\end{gather*}
 By solving the remaining equations for $M$, we find
\begin{gather*}
 M(x_1, x_2) = a^2 \beta_{13} \frac{x_2}{2 x_1^4}-2 \beta_{13} c \ln|x_1| \frac{x_2}{x_1^2}+ \beta_{13} (c-2 w) \frac{x_2}{x_1^2}.
\end{gather*}
However, this system is just a special case of \eqref{WSCd}, \eqref{BSCd} for $b=0$.

\subsection*{Acknowledgments}
This paper was supported by the Czech Science Foundation (Grant Agency of the Czech Republic), project 17-11805S.
This paper is dedicated to our son Flavio born just after the submission of the original manuscript.

\pdfbookmark[1]{References}{ref}
\LastPageEnding

\end{document}